%% file: reviewrelmatter.tex
\documentclass[preprint,preprintnumbers,amsmath,amssymb]{revtex4}
\usepackage{graphicx}
\newcommand{\beq}{\begin{equation}}
\newcommand{\eeq}{\end{equation}}
\newcommand{\beqa}{\begin{eqnarray}}
\newcommand{\eeqa}{\end{eqnarray}}
\newcommand{\nn}{\nonumber}
\newcommand{\Sigs}{\Sigma_{\mathrm s} }
\newcommand{\Sigv}{\Sigma_{\mathrm v} }
\newcommand{\Sigo}{\Sigma_{\mathrm o} }
\newcommand{\kf}{k_{\mathrm F} }
\newcommand{\kfj}{k_{\mathrm Fj} }

\newcommand{\kfp}{k_{\mathrm Fp} }
\newcommand{\bfgamma}{\mbox{\boldmath$\gamma$\unboldmath}}
\newcommand{\veck}{\textbf{k}}

\newcommand{\vecq}{\textbf{q}}

\newcommand{\Sigsn}{\Sigma_{{\mathrm s},n} }

\newcommand{\Sigon}{\Sigma_{{\mathrm o},n} }
\newcommand{\Sigsp}{\Sigma_{{\mathrm s},p} }

\newcommand{\Sigop}{\Sigma_{{\mathrm o},p} }
\newcommand{\Sigsi}{\Sigma_{{\mathrm s},i} }
\newcommand{\Sigvi}{\Sigma_{{\mathrm v},i} }
\newcommand{\Sigoi}{\Sigma_{{\mathrm o},i} }

\newcommand{\SigoDDRMFi}{\Sigma^{DDRMF}_{{\mathrm 0},i} }
\newcommand{\SigoDDRH}{\Sigma^{DDRH}_{\mathrm 0} }

\DeclareMathOperator{\Tr}{Tr}

\begin{document}

\title{Relativistic Effects in Nuclear Matter and Nuclei}

\author{Eric van Dalen}

\author{Herbert M\"uther}

\affiliation{ Institut f\"ur Theoretische Physik,
Universit\"at T\"ubingen, \\ Auf der Morgenstelle 14  
D-72076 T\"ubingen, Germany}

\begin{abstract}
The status of relativistic nuclear many-body calculations of nuclear systems to be built up in terms of protons and neutrons is reviewed.  In detail,
relativistic effects on several aspects of nuclear matter such as the effective mass, saturation mechanism, and the symmetry energy are considered. 
This review  will especially focus on isospin asymmetric issues, since these aspects are of high interest in astrophysical and nuclear structure studies. Furthermore, from the experimental side these aspects are experiencing an additional  boost from a new generation of radioactive beam facilities, e.g. the future GSI facility FAIR in Germany or SPIRAL2 at GANIL/France. Finally, the prospects of studying finite nuclei in microscopic calculations which are based on realistic $NN$ interactions by including relativistic effects in calculations of low momentum interactions are discussed.
\end{abstract}
\maketitle

\section{Introduction}
Nuclear systems are very intriguing many-particle systems, which have to be
treated by means of quantum theory. In this article we will consider nuclear
systems at low energies to be built up in terms of protons and neutrons, 
ignoring to a large extent all features which are related to the substructure of
the baryons. As compared to other areas for the application of quantum 
many-body or condensed matter theory, nuclear systems are of finite size. The
success of the Weiz\"acker mass formula in describing the binding energies of
nuclei shows that the underlying nucleon-nucleon ($NN$) forces are of short range
and that these binding energies result from an interplay between volume and
surface contributions. 

The success of this mass formula also allows to complement the nuclear systems
of finite size, i.e. the nuclei, by a theoretical construct, the system of
homogeneous infinite nuclear matter. Assuming that there is no Coulomb repulsion
between the nucleons one can construct isospin symmetric nuclear matter assuming
an infinite system of constant density with the same number of protons and
neutrons. The basic properties of this symmetric infinite matter can be deduced
from finite nuclei by extrapolation. The energy per nucleon should correspond to
the volume term in the Weiz\"acker mass formula, i.e. the energy per nucleon
of nuclei with identical number of protons and neutrons, ignoring surface and
Coulomb contributions. This energy per nucleon of about -16 MeV must be obtained
for the optimal density, i.e. the saturation density $\rho_0$, the density at
which the energy per nucleon is minimal. This saturation density can be deduced
from experimental data by extrapolating the baryon densities reached in the 
center of heavy nuclei. This leads to a value of about 0.16 nucleon/fm$^{-3}$.
So one first goal of nuclear many-body calculations is to reproduce these
empirical data of the saturation point for symmetric infinite nuclear matter.

However, infinite nuclear matter is not only a theoretical construct. The
collapse of stars after burning all the nuclear fuel in the fusion processes can
lead to a nuclear system with a homogeneous density over ranges which are so
large as compared to the typical distances between the nucleons, that surface
effects can be ignored. Since, however, these real systems of infinite nuclear
matter can not switch off the electro-magnetic interaction, the positive charges
of the protons must be compensated by a corresponding number of electrons or
muons with negative charge leading to a system in equilibrium with respect to
$\beta$-decay, which consists predominantly out of neutrons with a small
admixture (typically a few percent) of protons. Therefore these objects are
called neutron stars. It is one of the challenges of
nuclear many-body theory to predict the energy density of these infinite nuclear
systems with large proton-neutron asymmetries over a wide range of densities.
The equation of state resulting from such calculations is an important
ingredient for the simulation of neutron stars or astrophysical phenomena like 
supernovae.

Of special interest in this context is the study of the crust of such neutron
stars. At the low densities in the outer crust of neutron stars one can expect
the baryonic matter to be described as a lattice of ``normal'' nuclei, whereas
the baryonic matter with homogeneous density should be formed only at larger
densities (larger than about  50 percent of the saturation density of symmetric nuclear
matter) in the inner crust of the neutron star. The transitional region
between the isolated nuclei and the homogeneous matter is a very interesting
example for the phase transition of a quantum-liquid from the droplet phase to
the homogeneous phase. For this transitional region intriguing structures have
been predicted like quasi-nuclear bubbles, strings or layers embedded in a sea of
neutrons. Because of such structures this transitional region has been called
the pasta-phase of neutron stars~\cite{bv81,Montani:2004,Goegelein:2007}.

These introductory remarks shall demonstrate that the nuclear many-body problem
is indeed a very interesting topic in theoretical physics. It is of interest on
its own, because we want to understand the bulk properties of nuclei in terms of
the underlying $NN$ interaction. Of special interest is the study of exotic nuclei
with a large neutron excess and the extrapolation to the matter in neutron star,
since this provides the information for the simulation of astrophysical
processes. Theoretical studies of the nuclear many-body problem in terms of
nucleonic degrees of freedom are also of large
interest because they may finally exhibit the limitations of such nuclear structure
studies ignoring the sub-nucleonic degrees of freedom even in describing low
energy properties of nuclei.

Therefore a lot of effort has been made during the past decades to solve this
conventional nuclear many-body problem. In general one may distinguish between
so-called microscopic or \textit{ab initio} approaches on one hand and
more phenomenological approaches on the other hand. 

In what we will call the
phenomenological approach, one starts with a simple model for a $NN$ or even 3N
interaction and adjusts the parameter of this model to describe the saturation
point of symmetric nuclear matter and properties of a few selected nuclei. A
typical example of such a phenomenological interaction is the family of Skyrme
interactions~\cite{sk1,sk2,Skyrmest}. In this way one can describe rather
accurately properties of nuclei in detail. However, there may be a problem in
the predictions derived from these phenomenological studies for the systems at
large densities and proton-neutron asymmetries, which are relevant for the 
study of the astrophysical objects.

In the microscopic approach the parameters of the $NN$ interaction are adjusted to
describe the experimental data of $NN$ scattering including energies up to the
threshold for pion production and the properties of the deuteron. The $NN$
scattering data, in particular the change of sign in the S-wave phase shifts
suggest that such a realistic $NN$ interaction should contain repulsive components
of short range and attractive components with medium and long ranges. In fact,
some of the early models for a local $NN$ interaction considered a hard core
potential, with infinite repulsion for relative distances below $r_c \approx$
0.4 fm~\cite{hamada,reid}. Also modern versions for such a local $NN$ interaction
contain a considerable amount of cancellation between very repulsive components
of short range and attractive ones at larger ranges~\cite{wiringa:1984}.

But also other models, like the one-boson-exchange model~\cite{machl:89}, which
is based on the meson exchange theory for the $NN$ interaction and leads to a
non-local potential, contain rather strong components which are compensating
each other to a large extent.   

Employing such realistic $NN$ interaction with very strong components it turns
out, that perturbative approaches in terms of the $NN$ interaction do not lead to
reliable results. In fact, using such interactions within a mean field or
Hartree-Fock approach does not provide binding energy for nuclear
matter~\cite{muether:2000}. Therefore attempts have been made to sum up all ladder
diagrams and evaluate the energy of nuclear matter in terms of the $T$ matrix.
Hartree-Fock calculations in terms of this scattering matrix lead to a
sufficient amount of binding energy for symmetric nuclear matter but fail to
produce a saturation point: the energy per nucleon calculated in this way
decreases with increasing density and does not provide a minimum~\cite{machl:89}.

The situation is improved if one calculates the ladder diagrams accounting for
nuclear medium effects in the sense that the intermediate two-nucleon states are
restricted to the states above the Fermi momentum and by accounting for the
binding effects in two-nucleon propagator. This leads to the so-called Brueckner
G-matrix and the Brueckner-Hartree-Fock (BHF) approximation. 

The Pauli- and dispersive features included in the G-matrix lead to a minimum in
the calculated energy versus density plot, which means that they provide a
theoretical result for the saturation point of nuclear matter. A lot of effort
has been made to generate a realistic $NN$ interaction, which using the BHF
approximation also reproduces the experimental result for the saturation of
nuclear matter. However, all such calculations produced saturation points on the
so-called Coester band~\cite{Coester:70}. $NN$ interaction which are rather soft tend to
produce energies per nucleon which are close or even below the empirical value of 
-16 MeV but at a saturation density which about twice or even larger than the
empirical $\rho_0$. Other interactions, which are stiffer and contain larger tensor
components may predict the correct saturation density but yield a value for the
energy per nucleon, which is considerably large than the empirical value (-10 MeV
compared to -16 MeV). Anyway none of the interactions predicts a saturation point, 
which is close to the experimental result.

Attempts have been made to go beyond this BHF approximation. It is  one drawback of
this approach that particle-particle and hole-hole ladder terms are treated
differently. This implies that the BHF approximation is not symmetry conserving in
the sense of the Kadanoff and Baym~\cite{Kadan:62}, which implies e.g. that it does
not fulfill the Hugenholtz van Hove theorem. This has been improved by completing the
particle-particle ladders by corresponding hole-hole terms within the ladder
approximation of the self-consistent Greens function (SCGF) approach~\cite{frick:03}.
It turned out however, that the inclusion of the hole-hole terms does not lead to a
significant change for the predicted saturation properties of nuclear matter. This
supports the general strategy of the Brueckner Bethe hole line expansion.

The next step in the Brueckner hole-line expansion beyond the BHF approximation, is
to account for all three-particle ladders by solving the Bethe-Fadeev
equation~\cite{rajar:67}. Techniques and details, how to solve the Bethe-Fadeev
equation have been described by Day~\cite{day:81}, numerical calculations for
including the three-hole line terms have been presented by Song et al.~\cite{song:98}.
These studies show that the inclusion of three hole line terms provides the nice
feature that the results are getting essentially independent on the choice for the
particle spectrum, a choice which can not be decided within the the two hole line
approach. However, also the three hole line terms yield only a small correction  as
compared to the BHF approximation and therefore the predicted saturation points still
lead to a Coester band, which fails to come close to the empirical result.

A similar feature can also be observed in the calculation of bulk properties of
finite nuclei using the BHF approximation applied to realistic $NN$
interactions~\cite{muether:2000}. Such calculations tend to predict to little binding
energy and radii for the charge distribution, which are to small as compared to the
experimental findings.

The same is true for other many-body techniques, which have been used to evaluate the
properties of infinite nuclear matter or finite nuclei employing realistic $NN$
interactions within a non-relativistic framework. Such many-body techniques include
the coupled cluster or exponential S method~\cite{kuemmel:78,mihail:2000}, variational
methods allowing for correlated basis functions~\cite{bisconti:06} and Quantum Monte
Carlo Calculations~\cite{pieper:05}. Results close to the experimental data could only
be obtained after introducing empirical three nucleon forces.  

An alternative method, which is also based on realistic $NN$ interactions, has received
a lot of attention during the last few years: the use of low-momentum interaction 
$V_{\text{low-k}}$. The basic idea of $V_{\text{low-k}}$ is to separate the long
range or low-momentum components of  the $NN$ interaction,
which are constrained by the $NN$ scattering matrix below the pion threshold,
from the high-momentum components, which may strongly depend on the underlying model
of the $NN$ interaction. By introducing a cutoff $\Lambda$ in momentum
space, one separates the Hilbert space into a low-momentum and a high-momentum
part. The renormalization technique (see, e.g.~\cite{Bogner01,Lee80,Okubo54,Bogner08,Suzuki82}) 
determines an effective Hamiltonian, which treated within the low-momentum
space yields the same results than the original Hamiltonian for the corresponding
states. This renormalization can be achieved by means of the unitary model operator
approach~\cite{Lee80,Okubo54}. It is a nice feature of this $V_{\text{low-k}}$
interaction approach, that using a cutoff $\Lambda$ of 2 fm$^{-1}$, which corresponds
to the pion threshold in $NN$ scattering, one obtains a renormalized interaction, which
is essentially independent on the underlying realistic interaction, which was used to
generate $V_{\text{low-k}}$. Therefore one has got a unique $NN$ interaction model.

Another advantage of using this $V_{\text{low-k}}$ approach is that the strong short
range components, which are characteristic of a realistic $NN$ interaction have been
integrated out and a Hartree-Fock (HF) calculation is a reliable approximation for the
solution of the many-body problem. The price one has to pay however, is that such HF
calculations using $V_{\text{low-k}}$ interaction do not lead to a saturation point
when calculating the binding energy of infinite nuclear
matter~\cite{Bozek06,Kuckei03}. This is very similar to the results obtained in HF
calculations using the $NN$ $T$ matrix already discussed above~\cite{machl:89}. Indeed 
$V_{\text{low-k}}$ can be considered to be rather close to the $T$ matrix with the
only difference that the $NN$ states below the cutoff are excluded from the ladder
terms in $V_{\text{low-k}}$ as compared to $T$. Also in this case a possible way out
of the saturation problem is to include a phenomenological three-nucleon force.

All this demonstrates that three-nucleon forces seem to be necessary to obtain the
empirical saturation property of nuclear matter within a non-relativistic theory
based on realistic models of the $NN$ interaction.

On the other hand, however, relativistic calculations have been successful to derive the
empirical saturation point of symmetric nuclear matter from realistic meson exchange
models of the $NN$ interaction without the necessity to adjust appropriate
three-nucleon
forces.~\cite{brockmann:1990,anastasio:1983,serot:1986,terhaar:1987,weigel:1988,amorin:1992,sehn:1997,dejong:1998,gross:1999,schiller:2001,alonso:2003,vandalen:2004b,vandalen:2007} At first sight relativistic effects seem to be negligible
when dealing with the nuclear many-body problem, as the binding energies of nucleons are
very small as compared to the nucleon rest mass. The meson exchange model of the $NN$ 
interaction, however, predicts a strong cancellation of various components in the 
relativistic self-energy of nucleons in a nuclear medium. A strong attractive part 
(around -300 MeV), which transforms like a
scalar under a Lorentz transformation and mainly originates from the
exchange of the scalar $\sigma$ meson in realistic One-Boson-Exchange
(OBE) models of the $NN$ interaction, is partly compensated
by a repulsive component, which transforms like a time-like component of
a Lorentz vector and reflects the repulsion due to the $\omega$ meson
exchange. The effects of this two contributions cancel each other to a
large extent in calculating the single-particle energy of the nucleon,
which is indeed small (-40 MeV) on the scale of the mass of the nucleon.

Inserting the nucleon self-energy in a Dirac equation for the nucleon,
the strong scalar part of the
self-energy yields an enhancement of the small component in the
Dirac spinor of the nucleon in the medium as compared to the vacuum.
This modification of the Dirac spinors in the nuclear medium leads to
modified matrix elements for the OBE interaction. It is this density
dependence of the nucleon Dirac spinors and the resulting medium
dependence of the $NN$ interaction, which moves the saturation point
calculated for nuclear matter off the Coester band in such a way, that
the empirical values for the energy per nucleon and the saturation
density can be reproduced without the necessity to adjust any parameter
or to introduce many-nucleon forces. Reasonable saturation properties have
even been obtained within the framework of the $V_{\text{low-k}}$ approach, if
this change of the nucleon Dirac spinors in the nuclear medium is taken into
account~\cite{vandalen:2009a}. Therefor one may argue that the three-nucleon 
forces, which are required to achieve saturation within a non-relativistic
treatment, is an attempt to simulate the effects of these relativistic
features.

Although the relativistic many-body approach, the Dirac-Brueckner-Hartree-Fock
approach in particular, has been very successful in describing the saturation
property of nuclear matter, only a few attempts have been made to extend
the DBHF approximation also to the evaluation of finite
nuclei~\cite{muth:1987,fritz:1993,fritz:1994,ulrych:1999}. 
The consistent treatment of correlation 
and relativistic effects for finite systems is a rather involved problem. 

On the other hand, very extensive studies of finite nuclei have been made within
relativistic mean field approximation~\cite{vretenar:2005,ring:1996} using 
a parametrization which is adjusted to describe the experimental data of finite
nuclei. With the advance of the renormalization techniques leading to effective
low-momentum interactions $V_{\text{low-k}}$ the time may be appropriate to
consider the possible bridge between \textit{ab initio} relativistic calculations
based on $NN$ interactions fitted to the scattering data and these phenomenological
relativistic mean field calculations. 

We hope that this review on the status of relativistic nuclear many-body calculations
will help to built this bridge.

\section{Relativistic Many-Body Approaches}
During the past decades, much success has been achieved in nuclear physics by
relativistic many-body approaches. These approaches are the Dirac-Brueckner-Hartree-Fock
(DBHF) approaches and relativistic density functional theories. The DBHF approach is an
\textit{ab initio} approach, whereas relativistic density functional theory is a
phenomenological approach.

These relativistic Brueckner calculations are not straightforward and the approaches of
various groups
~\cite{brockmann:1990,terhaar:1987,weigel:1988,amorin:1992,sehn:1997,gross:1999,dejong:1998,gross:1999,schiller:2001,alonso:2003,vandalen:2004b,vandalen:2007}
are similar but differ in detail, depending on solution techniques and the particular
approximations made. Therefore, a general description of the relativistic Brueckner
approach will be given and the implications of the several approximation schemes will be
discussed.

One of the most outstanding relativistic density functional theories is the relativistic Hartree approach with the no-sea approximation, namely the relativistic mean-field (RMF) theory. Therefore, the DBHF approach is treated in Sec.~\ref{subsec:DBHF}, the RMF theory will be discussed in Sec.~\ref{subsec:RMF}, and the extensions to the RMF theory in Sec.~\ref{subsec:ermf}.

\subsection{DBHF approach}
\label{subsec:DBHF}
The main framework of the relativistic DBHF approach consists of a set of coupled equations, in which the
interaction of the nucleons, the propagation of the particles, and self-energies occur.
Firstly, the in-medium interaction of the nucleons is treated in the ladder
approximation of the relativistic Bethe-Salpeter (BS) equation
\beqa
T = V + i \int  V Q G G T,
\label{subsec:SM;eq:BS}
\eeqa
where $T$ denotes the $T$ matrix, $V$ the bare nucleon-nucleon, and $Q$ the Pauli operator, which
prevents scattering to occupied states. Furthermore, the Green's function $G$, which describes  the propagation of the dressed intermediate off-shell nucleons in Eq.~(\ref{subsec:SM;eq:BS}), fulfills the Dyson equation
\beqa
G=G_0+G_0\Sigma G,
\label{subsec:SM;eq:Dysoneq}
\eeqa
where $G_{0}$ denotes the free nucleon propagator and $\Sigma$ represents the self-energy.
This self-energy $\Sigma$ in the Hartree-Fock approximation is given by
\beqa
\Sigma = -i \int\limits_{F} (Tr[G T] - GT ).
\label{subsec:SM;eq:HFselfeq1}
\eeqa
Therefore, this coupled set of equations,
Eqs.~(\ref{subsec:SM;eq:BS})-(\ref{subsec:SM;eq:HFselfeq1}),
represents a self-consistency problem and has to be iterated until
convergence is reached.

The structure of the self-energy follows from translational and rotational invariance,
hermiticity, parity conservation, and time reversal invariance.
Therefore, the most general form of the Lorentz structure of the self-energy  is given by
\beqa
\Sigma = \Sigma_s - \gamma_\mu \Sigma^\mu
\label{subsec:DBHF;eq:selfour1}
\eeqa
with
\beqa
\Sigma^\mu = \Sigo u^\mu + \Sigv  \Delta^{\mu\nu}k_\nu,
\label{subsec:DBHF;eq:selfour2}
\eeqa
where the $\Sigs$, $\Sigo$, and $\Sigv$ components are Lorentz scalar
functions, which depend on the Lorentz invariants $k^2$, $k
\cdot  j$ and $j^2$, with $j_{\mu}$ the baryon current and $k_{\mu}$ the nucleon four-momentum.
Therefore, these Lorentz invariants can be expressed in terms of
$k_0$, $|\veck|$ and the Fermi momentum $\kf$.
Furthermore, the projector $\Delta^{\mu\nu}$ in Eq.~(\ref{subsec:DBHF;eq:selfour2}) is given by
$\Delta^{\mu\nu} = g^{\mu\nu} - u^\mu u^\nu$ with streaming velocity $u^\mu = j^\mu / \sqrt{j^2}$.
In the nuclear matter rest frame, $u^{\mu} = \delta^{\mu 0}$, the self-energy then has the simple form
\beqa
\Sigma(k,\kf)= \Sigs (k,\kf) -\gamma_0 \, \Sigo (k,\kf) +
\bfgamma  \cdot \textbf{k} \,\Sigv (k,\kf).
\label{subsec:SM;eq:self1}
\eeqa
These $\Sigs$, $\Sigo$, and $\Sigv$ components of the self-energy can be easily  calculated by taking the
respective traces~\cite{sehn:1997,horowitz:1987}
\beqa
\Sigs = \frac{1}{4} tr \left[ \Sigma \right],\quad
\Sigo = \frac{-1}{4} tr \left[ \gamma_0 \, \Sigma \right], \quad
\Sigv =  \frac{-1}{4|\veck|^2 }
tr \left[{\bfgamma}\cdot \veck \, \Sigma \right].
\label{subsec:SM;eq:trace}
\eeqa
\\ \indent The masses and momenta of the nucleons inside nuclear matter are modified by the nuclear medium
\beqa
m^*(k, \kf) = M + \Re e \Sigma_s(k, \kf), \quad k^{*\mu}=k^{\mu} + \Re e
\Sigma^{\mu}(k, \kf).
\label{subsec:SM;eq:dirac}
\eeqa
Written in terms of these effective quantities, the Dirac equation has the form
\begin{equation}
\left[ k^* \!\!\!\!\!\! / - m^* -i \, \Im m\, \Sigma \right] u(k) =0.
\label{dirac}
\end{equation}
The imaginary contribution to the self-energy is due to the possible decay of particle states above Fermi sea into hole status within the Fermi sea. To simplify  the self-consistency scheme this decay possibility is neglected and one
works in the quasi-particle approximation, i.e. $\Im m \Sigma = 0$. Furthermore, introducing the reduced effective mass,
\beqa
{\tilde m}^*(k,\kf) = m^*(k,\kf)/ \left( 1+\Sigv(k,\kf)\right),
\label{subsec:SM;eq:redquantity}
\eeqa
and the reduced kinetic momentum,
\beqa
{{\tilde k}^*}_\mu = k^*_\mu / \left( 1+\Sigv(k,\kf)\right),
\eeqa
the Dirac equation written in terms of these reduced effective masses and
momenta has the form
\beqa
 [ \gamma_\mu {\tilde k}^{*^\mu} - {\tilde m}^*(k,\kf)] u(k,\kf)=0.
\quad
\label{subsec:SM;eq:dirac2}
\eeqa
The solution of the Dirac equation in Eq.~(\ref{subsec:SM;eq:dirac2}) is
\beqa
u_\lambda (k,\kf)= \sqrt{ \frac{ {\tilde E}^*(\veck)+ {\tilde m}^*_F}
{2{\tilde m}^*_F}}
\left(
\begin{array}{c} 1 \\
\frac{2\lambda |\veck|}{{\tilde E}^*(\veck)+ {\tilde m}^*_F}
\end{array}
\right)
\chi_\lambda,
\label{spinor}
\eeqa
where ${\tilde E}^*(\veck)=\sqrt{\veck^2+{\tilde m}^{*2}_F}$ denotes the reduced effective energy
and $\chi_\lambda$ a two-component Pauli spinor with
$\lambda=\pm {\frac{1}{2}}$.
The normalization of the Dirac spinor is
thereby chosen as $\bar{u}_\lambda(k,\kf) u_\lambda(k,\kf)=1$.
From the Dirac equation in Eq.~(\ref{subsec:SM;eq:dirac2}) one derives the single-particle potential ${\hat U } = \gamma^0  \Sigma$. It can be obtained by calculating the expectation value of ${\hat U}$. Therefore, one sandwiches ${\hat U}$ between
the effective spinor basis (\ref{spinor}),
\begin{eqnarray}
   U(k) = \frac{\langle u(k)|\gamma^0  \Sigma | u(k)\rangle }
{\langle u(k)| u(k)\rangle } =
\frac{{\tilde m}^*}{{\tilde E}^*(\veck)}
\, \langle {\bar u(k)}| \Sigma | u(k)\rangle = \frac{{\tilde m}^*}{{\tilde E}^*(\veck)} {\tilde \Sigs} - {\tilde \Sigo}.
\label{upot1}
\end{eqnarray}

However, the results depend strongly on approximation schemes and techniques used in the DBHF approach.
In principal one can distinguish between two frequently used methods, the fit method versus the projection technique method.
The fit method is the more simple method; one avoids cumbersome projection techniques which are required using the trace formulas in Eq.~(\ref{subsec:SM;eq:trace}). This method was originally proposed by Brockmann and Machleidt~\cite{brockmann:1990}. In this approach,  the scalar and vector self-energy
components are directly extracted from the single particle potential U.
A fit to the single-particle potential $U$ in Eq.~(\ref{upot1}) delivers the density dependent components ${\tilde \Sigs}$ and ${\tilde \Sigo}$.
An attempt, which tries to extend
this method  and to extract momentum
dependent fields  by fitting procedures~\cite{lee:1997},   suffers from large uncertainties,
since one then tries to extract two functions out of one.
Therefore, only mean values for the self-energy components, where the explicit momentum dependence has already been averaged out, are relatively reliably obtained. However, the extrapolation to asymmetric matter already makes even this procedure
ambiguous for isospin asymmetric nuclear matter~\cite{schiller:2001}, since it introduces two new
parameters in order to fix the isovector dependencies of the
self-energy components.

The other method, the projection technique method, is rather complicated, but accurate. For example, they have been
used in Refs.~\cite{terhaar:1987,sehn:1997,gross:1999,vandalen:2004b,vandalen:2007}.
It requires the knowledge of
the Lorentz structure of the positive-energy-projected in-medium
on-shell $T$ matrix. The
$T$ matrix has to be projected onto covariant amplitudes. Therefore,
the scalar and vector components of the self-energies
can directly be determined from the projection onto Lorentz
invariant amplitudes. Ambiguities~\cite{nuppenau:1989}  arise due to the restriction to positive energy states,
since pseudoscalar ($ps$) and
pseudovector ($pv$) components
can not uniquely be disentangled for on-shell scattering.
However, these ambiguities can be minimized
by separating the leading order, i.e. the single-meson exchange, from
the full $T$ matrix. Therefore,  the contributions due to the single-$\pi$ and-$\eta$ exchange are given in the complete
$pv$ representation. For the remaining part of the $T$ matrix, the $ps$ representation
is chosen~\cite{vandalen:2004b,vandalen:2007}.

\subsection{Relativistic mean-field theory}
\label{subsec:RMF}
A Lagrangian density of interacting many-particle
system consisting of nucleons and mesons is the starting
point of a relativistic mean-field (RMF) theory. The different RMF models
differ from each other in the mesons and in the various couplings to the nucleon field included
in the Lagrangian density. The most simple RMF model is the linear $\sigma-\omega$ model,
which only includes linear coupling terms of the $\sigma$-meson and of the $\omega$-meson.
These models provide a relatively poor description of nuclear systems.
Therefore, nonlinear terms are added to the Lagrangian density to improve the description of isospin symmetric nuclear systems~\cite{walecka:1974} and isovector mesons are included in the models for the description of isospin asymmetric nuclear systems~\cite{liu:2002,margueron:2007}.

The Lagrangian density of the RMF theory presented here is taken as general as possible and includes as well the isoscalar mesons $\sigma$ and $\omega$ as the isovector mesons $\delta$ and $\rho$. The electromagnetic interaction is mediated by the photon $\gamma$.
Therefore, the Lagrangian density consists of four parts:
the free baryon Lagrangian density $\mathcal{L}_B$,
the free meson Lagrangian density $\mathcal{L}_M$,
the interaction Lagrangian density $\mathcal{L}_{\text{int}}$
and the nonlinear meson Lagrangian density $\mathcal{L}_{\text{nl}}$:
\begin{equation}\label{Lag_dens}
	\mathcal{L} = \mathcal{L}_B + \mathcal{L}_M + \mathcal{L}_{\text{int}} + \mathcal{L}_{\text{nl}},
\end{equation}
which takes on the explicit form
\begin{equation}
  \begin{split}
  \mathcal{L}_B =\,&  \bar{\Psi} ( \, i \gamma _\mu \partial^\mu - M ) \Psi,  \\
  \mathcal{L}_M =\,&  {\textstyle \frac{1}{2}} \sum_{\iota= \sigma, \delta}
			\Big( \partial_\mu \Phi_\iota \partial^\mu \Phi_\iota - m_\iota^2 \Phi_\iota^2 \Big)   \\
  		 &	- {\textstyle \frac{1}{2}} \sum_{\kappa = \omega, \rho, \gamma }
			\Big( \textstyle{ \frac{1}{2}} F_{(\kappa) \mu \nu}\, F_{(\kappa)}^{\mu \nu}
				- m_\kappa^2 A_{(\kappa)\mu} A_{(\kappa)}^{\mu} \Big),      \\
  \mathcal{L}_{\text{int}} =\,&	- g_\sigma\bar{\Psi}  \Phi_\sigma \Psi
                - g_\delta \bar{\Psi}  \boldsymbol{\tau} \boldsymbol{\Phi}_\delta \Psi
		 - g_\omega \bar{\Psi}  \gamma_\mu A_{(\omega)}^{ \mu } \Psi  \\ &
		- g_\rho \bar{\Psi}  \boldsymbol{\tau } \gamma_\mu  \boldsymbol{A}_{(\rho)}^{\mu } \Psi
  + \frac{f_\rho}{2M} \bar{\Psi}  \boldsymbol{\tau } \sigma_{\mu \nu}
		             [\partial^\nu\! \boldsymbol{A}_{(\rho)}^{\mu}] \Psi  \\ &
                - e \bar{\Psi}\gamma_\mu {\textstyle \frac{1}{2}}(1+ \tau_3 ) A_{(\gamma)}^{\mu} \Psi , \\
  \mathcal{L}_{\text{nl}}=\,& - \frac{1}{3}a\Phi_\sigma^{3} -\frac{1}{4}b\Phi_\sigma^{4} ,
  \end{split}
\end{equation}
with the field strength tensor
$F_{(\kappa)\mu \nu} = \partial_{\mu} A_{(\kappa)\nu} - \partial_\nu A_{(\kappa)\mu}$
for the vector mesons. In the above Lagrangian density the nucleon field is denoted by $\Psi$
and the nucleon rest mass by $M$.
The scalar meson fields are $ \Phi_\sigma$ and $ \boldsymbol{\Phi}_\delta $ and
the vector meson fields are $ A_{(\omega)} $ and $ \boldsymbol{A}_{(\rho)} $.
Furthermore, the bold symbols denote vectors in the isospin space acting between the two species of nucleons.
The mesons have rest masses $m_\sigma$, $m_\omega$, $m_\delta$, and $m_\rho$,
and couple to the nucleons with the strength of the coupling constants
$ g_\sigma$, $g_\delta$, $g_\omega$ , $g_\rho$, and $f_\rho$. The coupling constants of the first and second order nonlinear $\sigma$-meson self-interactions are $a$ and $b$, respectively.
Furthermore, the electromagnetic field $A_{(\gamma)}$ couples to the nucleons through the
electron charge $ e^2 = 4 \pi \alpha$ where $\alpha$
is the fine structure constant.

To obtain the field equations, we minimize the action
for variations of the fields $\phi_\kappa$
included in the Lagrangian density (eq. \ref{Lag_dens})
\begin{equation}
  \delta \int_{t_0}^{t_1} dt \int d^3x \,
      \mathcal{L}\big(\phi_\kappa(x), \partial_\mu \phi_\kappa(x), t \big) = 0,
\end{equation}
where $\phi_\kappa$ stands for the nucleon field, the meson fields
and the electromagnetic field. Finally the Euler--Lagrange field equations are obtained for each field $\phi_\kappa$
\begin{equation}
\frac{\partial}{\partial x^\mu}
\frac{\partial \mathcal{L} }{\partial(\partial_\mu \phi_\kappa)}
- \frac{\partial \mathcal{L} }{\partial \phi_\kappa} = 0.
\end{equation}
Evaluating the Euler--Lagrange field equations for each field $\phi_\kappa$,
we obtain a Dirac equation for the nucleons and
Klein-Gordon and Proca equations for the meson fields.
This Dirac equation for the nucleon field can be written as
\begin{equation}\label{Dirac_1}
 (i\gamma_\mu \partial^\mu - M -\Sigma )\, \Psi = 0
\end{equation}
with the self-energy $\Sigma$ generated by the interaction terms
\begin{equation}\label{self_en_1}
\begin{split}
  \Sigma =  \Big( & g_\sigma \Phi_\sigma
		+ g_\delta \boldsymbol{\tau} \boldsymbol{\Phi}_\delta
		+ g_\omega \gamma_\mu A_{(\omega)}^{ \mu }
		 + g_\rho \boldsymbol{\tau } \gamma_\mu  \boldsymbol{A}_{(\rho)}^{\mu }  \\
& -  \frac{f_\rho}{2M} \boldsymbol{\tau } \sigma_{\mu \nu}
		             [\partial^\nu\! \boldsymbol{A}_{(\rho)}^{\mu}]
		+ e\gamma_\mu {\textstyle \frac{1}{2}}(1+ \tau_3 ) A_{(\gamma)}^{\mu} \Big).
\end{split}
\end{equation}
Secondly, the equations for meson fields are given by
\begin{eqnarray}
( \Box + m_\sigma^2 ) \Phi_\sigma  + a{{\Phi_\sigma}^2}+ b{{\Phi_\sigma}^3}    \label{KG_sigma1}
	&=& - g_\sigma \bar{\Psi}  \Psi, \\\
( \Box + m_\delta^2 ) \boldsymbol{\Phi}_\delta
	&=& - g_\delta \bar{\Psi}  \boldsymbol{\tau} \Psi,  \\\
( \Box + m_\omega^2 ) A_{(\omega)\mu} &=& g_\omega \bar{\Psi} \gamma_\mu \Psi,  \\
( \Box + m_\rho^2 ) \boldsymbol{A}_{(\rho)\mu}
	&=& g_\rho \bar{\Psi} \boldsymbol{\tau} \gamma_\mu \Psi  + \partial^\nu {\textstyle\frac{f_\rho}{2M}} \bar{\Psi}
	      \boldsymbol{\tau} \sigma_{\mu \nu} \Psi ,  \\
 \Box  A_{(\gamma)\mu} &=& e \bar{\Psi}  \textstyle{ \frac{1}{2}} (1+\tau_3) \gamma_\mu \Psi.
\end{eqnarray}
Although these RMF models provide a reasonable description of nuclear matter and finite nuclei, medium effects
are still not satisfactorily treated.

\subsection{Extensions of the RMF theory.}
\label{subsec:ermf}

In this section, some extensions to the standard RMF theory will be discussed to obtain a more satisfactory treatment of the medium effects.
Therefore, RMF models were introduced, which allow for a density dependence of
meson-baryon vertices $g_\kappa$ and $f_\kappa$ with $\kappa$ a meson. Due to this density dependence of the coupling functions,
the nonlinear terms are not needed anymore and are omitted in these theories.
The density $\rho(\bar{\Psi}, \Psi)$ is obtained from the nucleon field $\Psi$. This introduced density dependence can improve the capability of the model significantly, in particular when it is based on microscopic many-body calculations. In  literature, a scalar density dependence
or a vector density dependence can be found. It turns out that the dependence on the zero component of the vector density,
the baryon density $\rho = \Tr( \bar{\Psi} \gamma_0 \Psi ) $,
is the most suitable one since it describes finite nuclei better
and has a natural connection to the vertices in microscopic many-body approaches like the DBHF calculations~\cite{hofmann:2001,gogelein:2008}.

These density dependent coupling functions can be determined by a fit either to experimental data or to the self-energies of the Dirac-Brueckner-Hartree-Fock approach for asymmetric nuclear matter. To be consistent with more involved theories for nuclear physics the second method
is preferred, since in contrast to the purely phenomenological relativistic mean-field model~\cite{ring:1996,niksic:2002},
the density dependence of the interaction has a microscopic foundation
when the density dependent RMF (DDRMF) approach is based on microscopic many-body calculations. This should give one confidence when the model is used for extreme cases such as neutron-rich nuclei. Therefore, the relation between the the DDRMF theory and DBHF calculations will be discussed in Sec.~\ref{sec:nuclei}.

As a consequence of this density dependence of meson-baryon vertices, we have to vary the Lagrangian density by
\begin{equation}
  \frac{\delta \mathcal{L}}{\delta \bar{\Psi}}
	= \frac{\partial \mathcal{L}}{\partial \bar{\Psi}}
	+ \frac{\partial \mathcal{L}}{\partial \rho} \frac{\delta \rho}{\delta \bar{\Psi}},
\end{equation}
where the second expression generates the so-called rearrangement
contribution $\Sigma^{(r)}$ to the self-energies of the nucleon field.
This rearrangement
contribution $\Sigma^{(r)}$ has to be added to the Dirac equation (\ref{Dirac_1})
\begin{equation}
  (i\gamma_\mu \partial^\mu - M -\Sigma )\, \Psi = 0
   \longrightarrow
   \big(i\gamma_\mu \partial^\mu - M - ( \Sigma + \Sigma^{(r)} \gamma_0 )\big) \, \Psi = 0,
\end{equation}
where the rearrangement self-energy contribution $\Sigma^{(r)}$ reads
\begin{equation}\label{self_en_rearr_1}
\begin{split}
  \Sigma^{(r)} = \Big(
	& \frac{\partial g_\sigma}{\partial \rho} \bar{\Psi} \Phi_\sigma \Psi
	 + \frac{\partial g_\delta}{\partial \rho} \bar{\Psi} \boldsymbol{\tau} \boldsymbol{\Phi}_\delta \Psi
	 + \frac{\partial g_\omega}{\partial \rho} \bar{\Psi} \gamma_\mu A_{(\omega)}^{ \mu } \Psi \\
	 & + \frac{\partial g_\rho}{\partial \rho}
	      \bar{\Psi} \boldsymbol{\tau } \gamma_\mu  \boldsymbol{A}_{(\rho)}^{\mu } \Psi
- \frac{1}{2M}\frac{f_\rho}{\partial \rho} \bar{\Psi} \boldsymbol{\tau } \sigma_{\mu \nu}
		             [\partial^\nu\! \boldsymbol{A}_{(\rho)}^{\mu}]  \Psi
            \Big).
\end{split}
\end{equation}
These rearrangement contributions should be taken into account and are essential
to provide a symmetry conserving approach, which implies that energy-momentum 
conservation and thermodynamic consistency like the Hugenholtz - van Hove
theorem are satisfied~\cite{Kadan:62,fuchs:1995}. 

Note that such rearrangement terms are not present in the microscopic DBHF 
approximation. That is why the DBHF as well as its not-relativistic
counterpart, the Brueckner-Hartree-Fock (BHF) approximation, are not symmetry
conserving. An extension of BHF, which accounts for particle-particle as well as
hole-hole ladders, would be required to achieve this
consistency~\cite{frick:03}. This implies that the definition of the
nucleon-selfenergy in (\ref{subsec:SM;eq:HFselfeq1}) must be extended by terms
of higher order in the hole-line expansion to obtain such a number conserving
approach~\cite{Kuo:1980}.  

Another extension of the RMF theory is the inclusion of Fock terms to obtain a relativistic Hartree-Fock (RHF) theory. In this approach, an additional meson, the $\pi$-meson, is usually included. This means that the Langrangian density has to be extended,
\begin{equation}
\mathcal{L}_{\pi}= \frac{1}{2}
			\Big( \partial_\mu \Phi_\pi \partial^\mu \Phi_\pi - m_\pi^2 \Phi_\pi^2 \Big)  -\frac{f_\pi }{m_\pi} \bar{\Psi}  \boldsymbol{\tau} \gamma_5 \gamma_\mu
			[\partial^\mu \boldsymbol{\Phi}_\pi]  \Psi,
\end{equation}
and consequently all equations derived from it. The presence of a $\pi$ field is a typical feature of the RHF theory compared to the RMF theory, since this $\pi$ field only contributes to the Fock terms. Therefore, it is omitted in the RMF theory.

\section{Effective Mass}
The introduction of an effective mass is a common concept in nuclear physics.
It is used to characterize the
quasi-particle properties of a particle inside a strongly interacting
medium. Furthermore, it is a well-known fact that the effective nucleon mass
in nuclear matter or finite nuclei deviates substantially from
its vacuum value~\cite{brown:1963,jeukenne:1976,mahaux:1985}.
However, the phrase of an effective nucleon mass  has been used to denote
different quantities: the nonrelativistic effective mass $m^*_{NR}$ and the
relativistic Dirac mass $m^*_{D}$.
Although these quantities are related, they are based on completely different physical concepts.

\begin{figure}[!t]
\begin{center}
\includegraphics[width=0.9\textwidth] {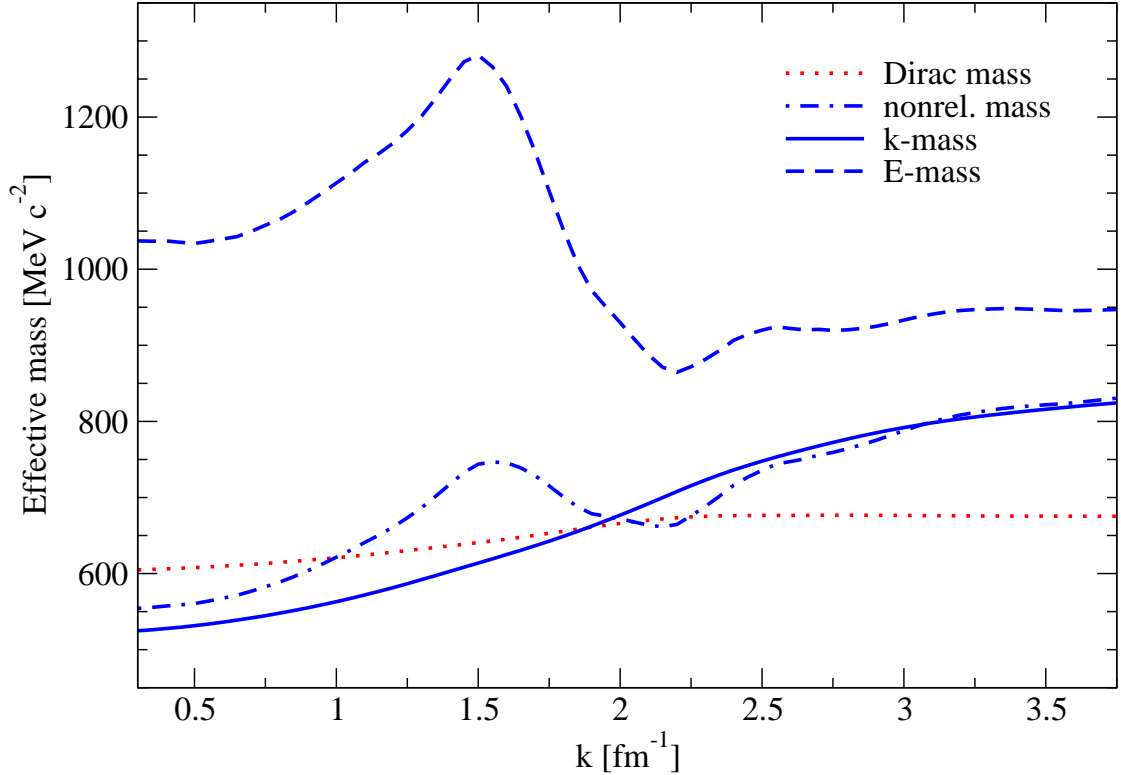}
\caption{The Dirac mass (dotted line), nonrelativistic masss $m^*_{NR}$ (dashed-dotted lines), $k$-mass $m^\ast_k(k)$ (solid lines), $E$-mass $m^\ast_E(k)$ (dashed lines)  in isospin symmetric nuclear matter as a function of the
momentum $k=|\veck|$ obtained from DBHF calculations at a fixed nuclear density of
$\rho = 0.181 \quad \textrm{fm}^{-3}$.
\label{fig:momdepmass}}
\end{center}
\end{figure}

The Dirac mass is a genuine relativistic quantity and can only be
determined from relativistic approaches. It is defined through the scalar part
of the nucleon self-energy in the
Dirac field equation which is absorbed into the effective mass
\beqa
m^*_D(k, \kf) = M + \Re  \Sigma_s(k, \kf),
\label{subsec:SM;eq:dirac3}
\eeqa
where $\Sigs$ is the scalar part of the nucleon self-energy.The Dirac mass accounts for medium effects through the scalar part of the self-energy and is a smooth function of the momentum as shown in Fig.~\ref{fig:momdepmass} for a DBHF calculation~\cite{vandalen:2010}.

On the other hand, the nonrelativistic mass $m^*_{NR}$ parameterizes
the momentum dependence of the single particle potential. Therefore,  it  can be
determined from both, as well relativistic as nonrelativistic
approaches. The nonrelativistic mass is defined as
\beqa
m^*_{NR} = |\veck| [dE/d|\veck|]^{-1},
\label{Landau1}
\eeqa
where $E$ is the quasi-particle's energy and $\veck$ its momentum. When
evaluated at $k=k_F$, Eq.(\ref{Landau1}) yields the Landau mass.
In the quasi-particle approximation, i.e. the zero width limit of the in-medium
spectral function, the energy $E$ of the quasi-particle and its momentum $\veck$
are connected by the dispersion relation
\beqa
E= \frac{\veck^2}{2 M} + \Re U.
\label{Energy1}
\eeqa
Hence, the equations (\ref{Landau1}) and (\ref{Energy1}) yield the
following expression for the nonrelativistic effective  mass
\beqa
m^*_{NR} = \left[\frac{1}{M}
+  \frac{1}{|\veck|} \frac{d}{ d |\veck|}\Re U (|\veck|, \omega=E)\right]^{-1}.
\label{mlandau}
\eeqa
Since the nonrelativistic effective mass parameterizes the momentum
dependence of the single particle potential, it also is a
measure of the nonlocality of the single particle potential $U$.

These nonlocalities of $U$ can be due to nonlocalities in space or in
time. The spatial nonlocalities result in a momentum dependence and are characterized by the
so-called k-mass, which is obtained from the
derivative of $U$ with respect to the momentum at {\it fixed} energy,
\begin{equation}
  \frac{m_k(k)}{M} = \left[ 1 + \frac{M}{k}
    \frac{\partial U ( k, \omega) }{\partial k} \right]^{-1}.
\end{equation}
These spatial nonlocalities of $U$ are mainly generated by exchange Fock terms~\cite{frick:2002,hassaneen:2004} and the resulting k-mass is a smooth function of the momentum
as shown in Fig.~\ref{fig:momdepmass} for a DBHF calculation~\cite{vandalen:2010}.

On the other hand, nonlocalities in time result in an energy dependence and are expressed in terms of the
E-mass, which is given by the
derivative of $U$ with respect to the energy at {\it fixed}
momentum,
\begin{equation}
  \frac{m_E(\omega)}{M} = \left[ 1 -
    \frac{\partial U ( k, \omega) }{\partial \omega} \right].
\end{equation}
The nonlocalities in time are generated e.g. by Brueckner ladder
correlations due to the scattering to intermediate off-shell
states. These correlations mainly have a short-range character
and generate a strong momentum  dependence with a characteristic
enhancement of the E-mass slightly above the Fermi surface~\cite{mahaux:1985,frick:2002,hassaneen:2004,jaminon:1989}.

The effective nonrelativistic
mass defined by Eqs. (\ref{Landau1}) and (\ref{mlandau}) is given
by the product of the k-mass and the E-mass~\cite{jaminon:1989},
\begin{equation}
  \frac{m^\ast(k)}{M} = \frac{m^\ast_k(k)}{M} \frac{m^\ast_E(\omega =
  E)}{M}.\label{eq:mstartot}
\end{equation}
Therefore, it
contains nonlocalities as well in space  as in time and
should show a  typical peak structure around $\kf$.
This narrow enhancement near the Fermi surface reflects  - as a model independent result - the increase
of the level density due to the vanishing imaginary part of the
optical potential at $\kf$, which for example is seen in shell model
calculations~\cite{jeukenne:1976,mahaux:1985,jaminon:1989}. In
order to reproduce this behavior, one should, however,
account for correlations beyond mean-field or Hartree-Fock.

The enhancement of the effective mass $m^\ast$ due to the
effective $E$-mass in Eq.(\ref{eq:mstartot}) is not strong enough to compensate for
the effects of the $k$-mass. Therefore, the final effective mass is below the
bare mass $M$ as can be seen in Fig.~\ref{fig:momdepmass}.

\begin{figure}[!t]
\begin{center}
\includegraphics[width=0.8\textwidth] {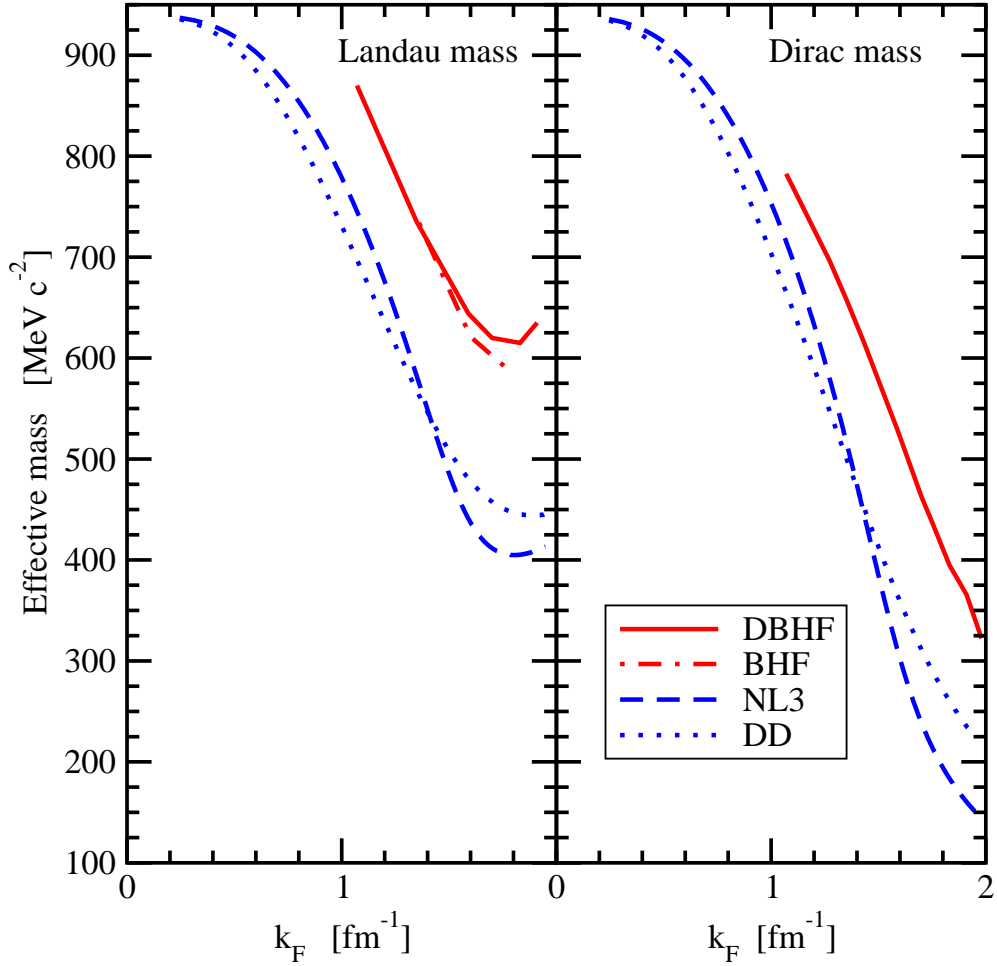}
\caption{The effective nonrelativistic (Landau) and Dirac mass in isospin symmetric nuclear matter
at $k=|\veck|=\kf$ as a function of the
Fermi momentum $\kf$ for various relativistic models are plotted. In addition, the Landau mass of a nonrelativistic
BHF calculation is shown.
\label{fig:mstar}}
\end{center}
\end{figure}
Another aspect is the density dependence of these two different masses. The Dirac mass of
relativistic models decreases continously with increasing Fermi momentum~$\kf$ as shown
in Fig.~\ref{fig:mstar}. The plotted results in Fig.~\ref{fig:mstar} are from a
relativistic DBHF approach~\cite{vandalen:2005a,vandalen:2005b} based on the Bonn A 
interaction and from RMF calculations of the NL3 and DD model. The NL3 model is a
nonlinear  parameterization~\cite{lalazissis:1997} that is widely used in RMF
calculations. It contains the $\sigma$-meson, the $\omega$-meson, the $\rho$-meson
without its tensor coupling, the electromagnetic field, and nonlinear self-interactions
of the $\sigma$ field. Compared to the Lagrangian density of the NL3 model, the DD model
is based on a Lagrangian density without nonlinear self-interactions of the $\sigma$
field, however, with density dependent meson-nucleon coupling vertices~\cite{typel:2005}.
Initially, the Landau  mass also decreases with increasing Fermi momentum $\kf$ like the
Dirac mass. However, it starts to rise again at high values of the Fermi momentum $\kf$.
In addition to the results from relativistic approaches, the Landau mass of a
nonrelativistic BHF calculation, which is also based on the Bonn A  interaction, is
plotted in Fig.~\ref{fig:mstar}. The agreement between the calculated Landau mass from
the nonrelativistic and the relativistic Brueckner approach is quite good. Therefore, the
often discussed difference between effective masses obtained in the various approaches is
mainly due to different definitions, i.e. nonrelativistic mass versus Dirac mass.
Although the RMF models qualitatively show the same behavior as the Brueckner approaches,
their masses are lower compared to the ones in the Brueckner approaches.

Another interesting issue is the proton-neutron mass splitting in
isospin asymmetric nuclear matter. This issue will be of relevance in the study of drip-line nuclei.
Furthermore, it can be expected to have important effects on transport properties, like collective flows, of dense isospin asymmetric nuclear matter that will be reached in heavy ion experiments.

\begin{figure}[!t]
\begin{center}
\includegraphics[width=0.9\textwidth] {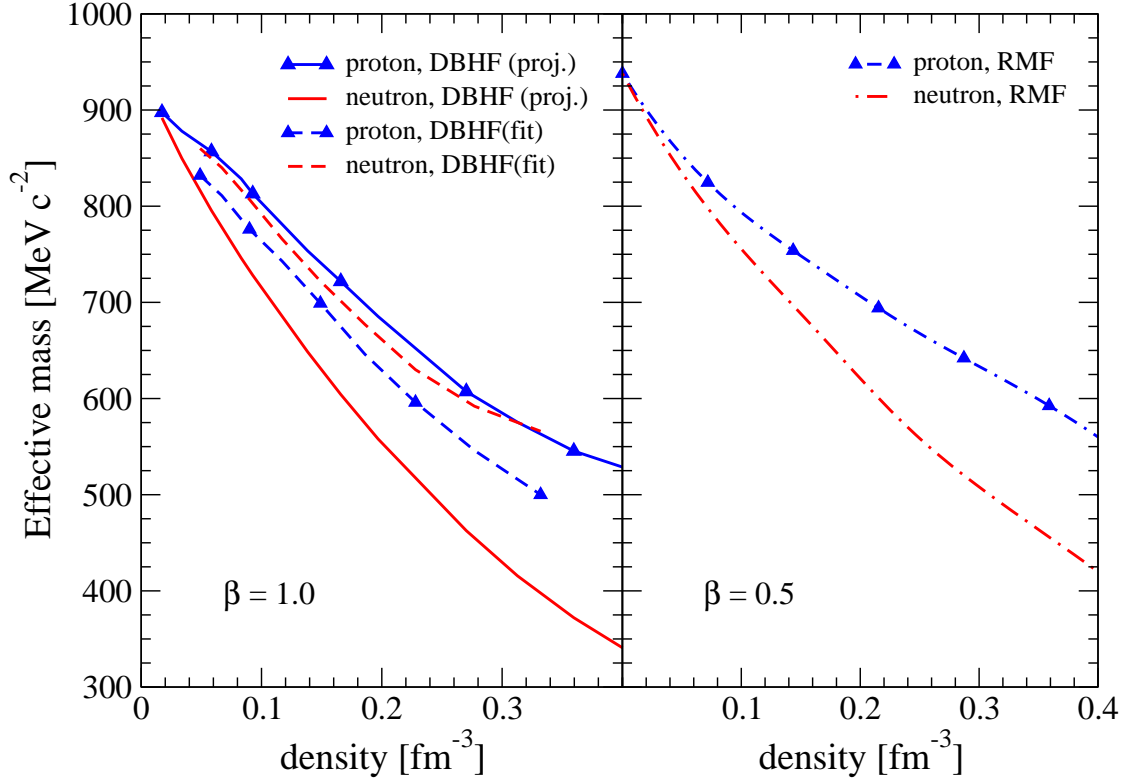}
\caption{Left panel: Results obtained from different DBHF calculations for the Dirac neutron and proton effective masses
as a function of the density $\rho$ for a value of the asymmetry parameter $\beta=\frac{\rho_n-\rho_p}{\rho}=1$. One of the DBHF calculations is based
on projection techniques and  the other calculation is based on the fit method.
Right panel: The Dirac neutron and proton effective masses of a RMF theory, which includes the isovector $\delta$-meson, are shown.
\label{fig:Diracsplit}}
\end{center}
\end{figure}

The DBHF calculations
based on projection techniques predict a Dirac mass splitting of
$m^*_{D,n} < m^*_{D,p}$ in neutron-rich nuclear
matter~\cite{dejong:1998,schiller:2001,vandalen:2004b,vandalen:2007,vandalen:2005a,vandalen:2005b}, as can be observed  in Fig.~\ref{fig:Diracsplit} for such a DBHF calculation~\cite{vandalen:2007}.
%
%
In contrast, the
DBHF calculations based on the fit method yield a mass splitting of $m^*_{D,n} > m^*_{D,p}$~\cite{schiller:2001,alonso:2003,sammarruca:2005} as shown in the left panel of Fig.~\ref{fig:Diracsplit} for such a calculation~\cite{alonso:2003}. In the fit method, momentum independent self-energy components are obtained, since the
explicit momentum dependence has already been averaged out. Hence, only mean values for the self-energy components are obtained.  This method is relatively reliable in isospin
symmetric nuclear matter. However, the extrapolation to isospin asymmetric matter introduces two new
parameters in order to fix the isovector dependencies of the
self-energy components. This introduction of two new parameters makes the fit procedure
ambiguous in isospin asymmetric nuclear matter~\cite{schiller:2001,ulrych:1997}.
Furthermore, RMF theories with the isovector $\rho$- and $\delta$-mesons included predict
a Dirac mass splitting of
$m^*_{D,n} < m^*_{D,p}$~\cite{liu:2002,baran:2005} as you can see in the right panel of Fig.~\ref{fig:Diracsplit} for such a RMF calculation~\cite{liu:2002}. This behavior of the neutron-proton mass splitting is  in agreement with results from the DBHF calculations
based on projection techniques. When only the $\rho$-meson is included, the RMF theory predicts equal masses, $m^*_{D,n}= m^*_{D,p}$. Hence,
the $\delta$-meson is responsible for the mass splitting in the RMF theory.

\begin{figure}
\begin{center}
\mbox{
\includegraphics[width=0.8\textwidth]{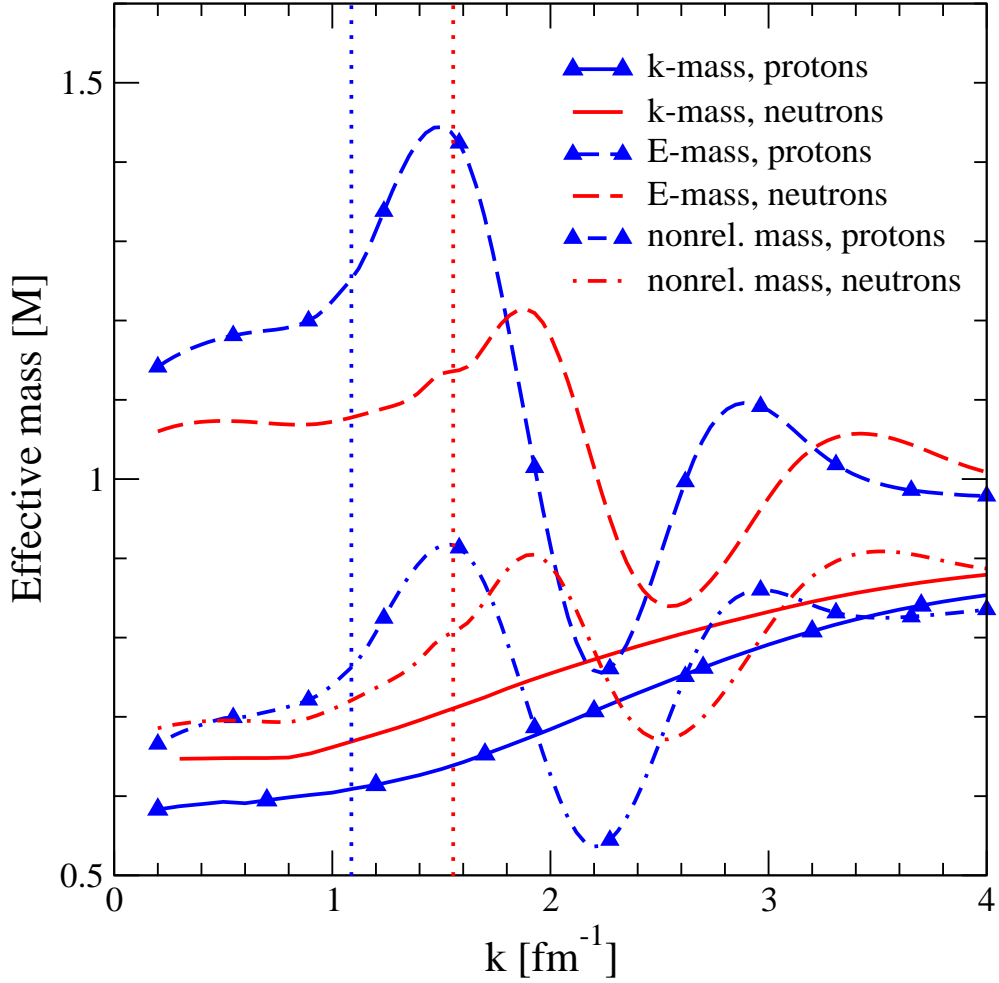}}
\end{center}
\caption{\label{fig:MeffBHFreview}
Effective $k$-mass $m^\ast_k(k)$ (solid lines), effective $E$-mass $m^\ast_E(k)$ (dashed lines) and effective nonrelativistic masss $m^*_{NR}$ (dashed-dotted lines) for neutrons and protons (lines with symbol) as obtained from the BHF calculations for a value
of the asymmetry parameter $\beta=0.5$ at fixed nuclear density
$\rho = 0.17 \quad \textrm{fm}^{-3}$. The
Fermi momenta for protons and neutrons are indicated by vertical dotted lines.}
\end{figure}

For completeness, attention will now be given to the nonrelativistic effective mass, which is obtained in Fig.~\ref{fig:MeffBHFreview} from a BHF calculation~\cite{gogelein:2009} by taking the product of the effective $k$-mass and the effective $E$-mass as done in Eq.~(\ref{eq:mstartot}).
The effective $k$-mass is larger for the neutrons than for the protons, whereas the
values for the effective $E$-mass are larger for the protons than for the neutrons in
neutron-rich nuclear matter. Since the difference between the effective $E$-mass of the
neutron and the proton is not large enough to compensate for the isospin asymmetric
nuclear matter effects of the $k$-mass, the nonrelativistic proton mass is in general
smaller than the one of the neutron, in particular at $k=|\veck|=\kf$. This result for
the  proton-neutron mass splitting  is a general feature of BHF
calculations~\cite{frick:2002,hassaneen:2004,zuo:1999,zuo:2005}. In contrast, the RMF
theorie and the DBHF calculations  based on on projection techniques predict an opposite
behavior for the Dirac mass spliting.

\begin{figure}[!t]
\begin{center}
\includegraphics[width=0.9\textwidth] {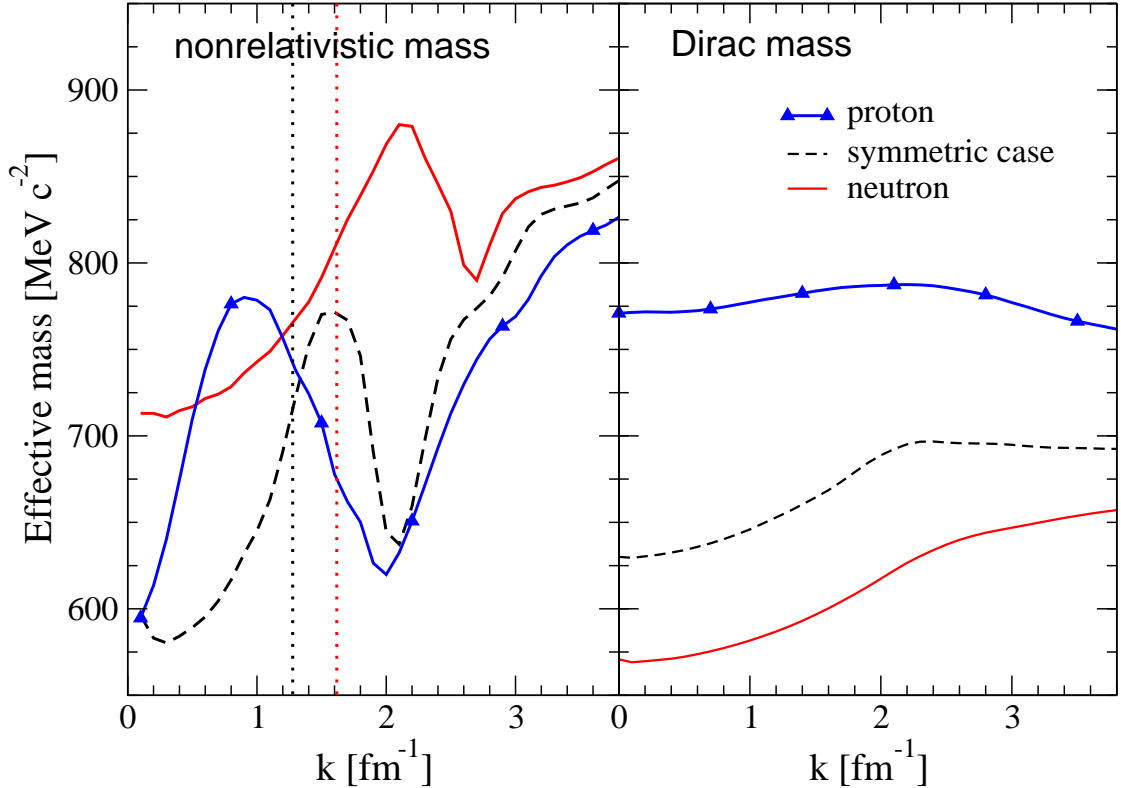}
\caption{Neutron and proton effective mass obtained from a DBHF calculation based on projection
techniques are plotted as a function of the
momentum $k=|\veck|$ for a value
of the asymmetry parameter $\beta=1$ at fixed nuclear density
$\rho = 0.166 \quad \textrm{fm}^{-3}$. In addition,
the effective mass in symmetric nuclear matter is given and the neutron
Fermi momenta  are indicated by vertical dotted lines for isospin symmetric nuclear matter and for pure neutron matter.
\label{fig:asplit}}
\end{center}
\end{figure}

As already mentioned before, the nonrelativistic mass  $m^*_{NR}$ can be
determined from both, as well relativistic  as nonrelativistic
approaches. The nonrelativistic mass derived from the RMF theory shows the same behavior as its
Dirac mass, namely  $m^*_{NR,n} < m^*_{NR,p}$~\cite{baran:2005}. The nonrelativistic mass derived from the DBHF approach based on projection techniques shows a nonrelativistic mass splitting of $m^*_{NR,n} > m^*_{NR,p}$ except around the peak slightly above the proton
Fermi momentum $\kfp$. This behavior is opposite to its Dirac mass splitting of $m^*_{D,n} <m^*_{D,p}$, as can be seen in
Fig.~\ref{fig:asplit} for a DBHF calculation based on projection
techniques~\cite{vandalen:2004b,vandalen:2005a,vandalen:2005b}.
%
%
However, it is in agreement with
results from nonrelativistic BHF calculations~\cite{frick:2002,hassaneen:2004,zuo:1999}. Since these masses are based on completely different physical concepts, the opposite behavior between the Dirac mass splitting and the nonrelativistic mass
splitting is not surprising. On the one hand,  the nonrelativistic mass defined by Eq.~(\ref{mlandau}) parameterizes
the momentum dependence of the single particle potential and is the result of a quadratic parameterization
of the single particle spectrum.  On the other hand, the relativistic Dirac mass
is defined by Eq.~(\ref{subsec:SM;eq:dirac3}) through the scalar part of the nucleon self-energy in the
Dirac field equation which is absorbed into the effective mass.

\section{Saturation}
\label{sec:sat}
In this section the saturation mechanism and properties of nuclear matter in relativistic many-body approaches
will be discussed. This saturation mechanism is quite different compared to nonrelativistic theories.
In contrast to nonrelativistic theories, one has large attractive scalar fields and repulsive vector fields of several hundred MeV. Saturation occurs through the interplay between these large attractive and repulsive fields. The small nuclear binding energy arises from the cancellation between the large attractive scalar field and the large repulsive vector field.

In RMF theory the large attractive scalar field $\Sigs$ is generated by the
$\sigma$-meson, and the repulsive vector field $\Sigo$ originates from the
$\omega$-meson~\cite{serot:1986,ring:1996}. In these relativistic approaches the vector field  grows linearly with density whereas
the scalar field saturates at large densities which leads finally to saturation.
Correlation effects and Fock exchange terms are not present in these RMF models. However, this does not mean that the saturation mechanism is dominated by the mean-field or Hartree contribution
and that exchange terms and higher correlations play only a minor role. In microscopic many-body approaches such as BHF and DBHF calculations based on realistic interactions, correlations and Fock exchange terms play an important role in the saturation mechanism. Therefore, RMF models which use the coupling strength values of the realistic interaction
can qualitatively reproduce saturation, but completely fails in a quantitative description. Therefore, the parameters
of RMF models have to be obtained from a fit to properties of finite nuclei and nuclear matter or are determined by a fit to the self-energies of a microscopic many-body calculation such as the DBHF approach.

\begin{figure}[!t]
\begin{center}
\includegraphics[width=0.9\textwidth] {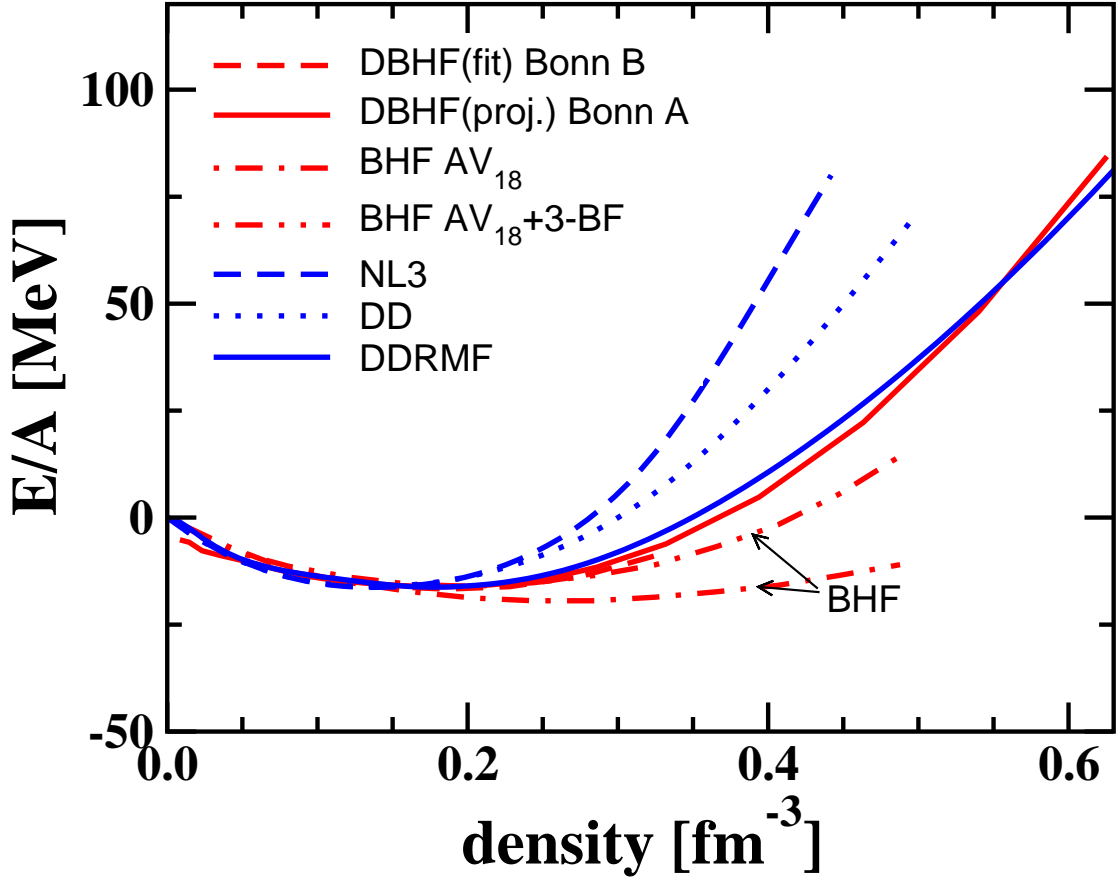}
\caption{Comparison of several EoSs from relativistic many-body approaches.
The relativistic ab initio approaches are a DBHF using the fit method and one using projection techniques. The relativistic mean-field models presented are the nonlinear NL3 model, the DD model, and the DDRMF model. In addition, a nonrelativistic BHF is shown with three-body force and without three-body force.
\label{fig:eos_DB}}
\end{center}
\end{figure}

The binding energy results of some relativistic microscopic many-body approaches, a DBHF calculation based on projections techniques~\cite{gross:1999,vandalen:2004b,vandalen:2007}  and one based on the fit method~\cite{alonso:2003},  are depicted in Fig.~\ref{fig:eos_DB}. The saturation behavior of three different RMF theories are shown in Fig.~\ref{fig:eos_DB}. They are the NL3~\cite{lalazissis:1997}, DD~\cite{typel:2005}, and DDRMF~\cite{gogelein:2008} parameterizations. The parameters of the NL3 and the density dependent coupling functions of the DD model were obtained from a fit to properties of finite nuclei. The parameters
of the DDRMF model are determined by a fit to the self-energies of a DBHF
approach  with a slight modification at saturation density to better reproduce
properties of finite nuclei. In addition, nonrelativistic BHF calculations from
Ref.~\cite{zuo:2002} based on the local Argonne $V_{18}$~\cite{wiringa:1995} are shown with and without three-body force. The included microscopic three-body force is based on a meson exchange model~\cite{zuo:2002}. The curves of all the microscopic and RMF methods are very similar below a density of approximately 0.2 fm$^{-3}$. This fact indicates that these are the density ranges which are at present reasonably well controlled by state-of-the-art many-body calculations. At higher densities noticeable differences occur.

Another observation that becomes evident from fig.~\ref{fig:eos_DB}, by comparing microscopic calculations to RMF calculations fitted to finite nuclei is that in nuclear matter the BHF and DBHF calculations lead to more binding than the RMF calculations. The prediction of a soft EoS is the general outcome of a microscopic  many-body calculation. Recent Quantum Monte Carlo calculations for symmetric nuclear matter~\cite{gandolfi:2006} show the same tendency. It should be noticed that this observation is supported by corresponding observables extracted from heavy ion reactions, where supranormal densities up to about three times saturation density are probed. Heavy ion data for tranverse flow~\cite{stoicea:2004} or from kaon production~\cite{sturm:2001} support the picture of a soft EoS in symmetric nuclear matter. The compression moduli of the RMF models, which are presented in Table~\ref{table:NM_Prop}, are too high compared to the experimental value of $K=225\pm15$ MeV, except for the DD model where the nuclear incompressibility was fixed to $K=240$ MeV. In general, RMF fits to finite nuclei
require a relatively high compression modulus K of about 300 MeV at saturation density~\cite{ring:1996}. However, at higher densities the DDRMF model has the softest EoS, since this special  procedure for the DD model can not avoid that at high density the EoS becomes very stiff.

\begin{table}
\begin{center}
\begin{tabular}{|ll|ccccc|}
\hline
	&& NL3~\cite{lalazissis:1997}  &  DD~\cite{typel:2005}  & DDRMF~\cite{gogelein:2008}  & DBHF~\cite{alonso:2003} & DBHF~\cite{vandalen:2004b,vandalen:2007}\\
        &&                              &                        &                              &(fit)                     & (proj.)\\
\hline
$\rho_0$ &       [fm$^{-3}]$ & 0.1482 &  0.1487  & 0.178  & $\approx$ 0.19 & 0.181 \\
$E/A$   &[MeV]	     & -16.24 &  -16.021 & -16.25 & -16.7        & -16.15 \\
$K$      &       [MeV]       & 271.5  &  240     & 337    & 233          & 230       \\
$a_S$   & [MeV]       & 37.4   &  31.6    & 32.1   & $\approx$ 30   & 34.36 \\
\hline
\end{tabular}
\end{center}
\caption{\label{table:NM_Prop}
Saturation properties of nuclear matter are compared for the NL3 model, the DD model, the DDRMF model, a DBHF approach based on the fit method, and a DBHF approach based on projection techniques.
The quantities listed include the saturation density $\rho_0$, the
binding energy $E/A$ at saturation density, the compressibility modulus $K$ and the
symmetry energy $a_S$ at saturation density.}
\end{table}

\begin{figure}[!t]
\begin{center}
\includegraphics[width=0.9\textwidth] {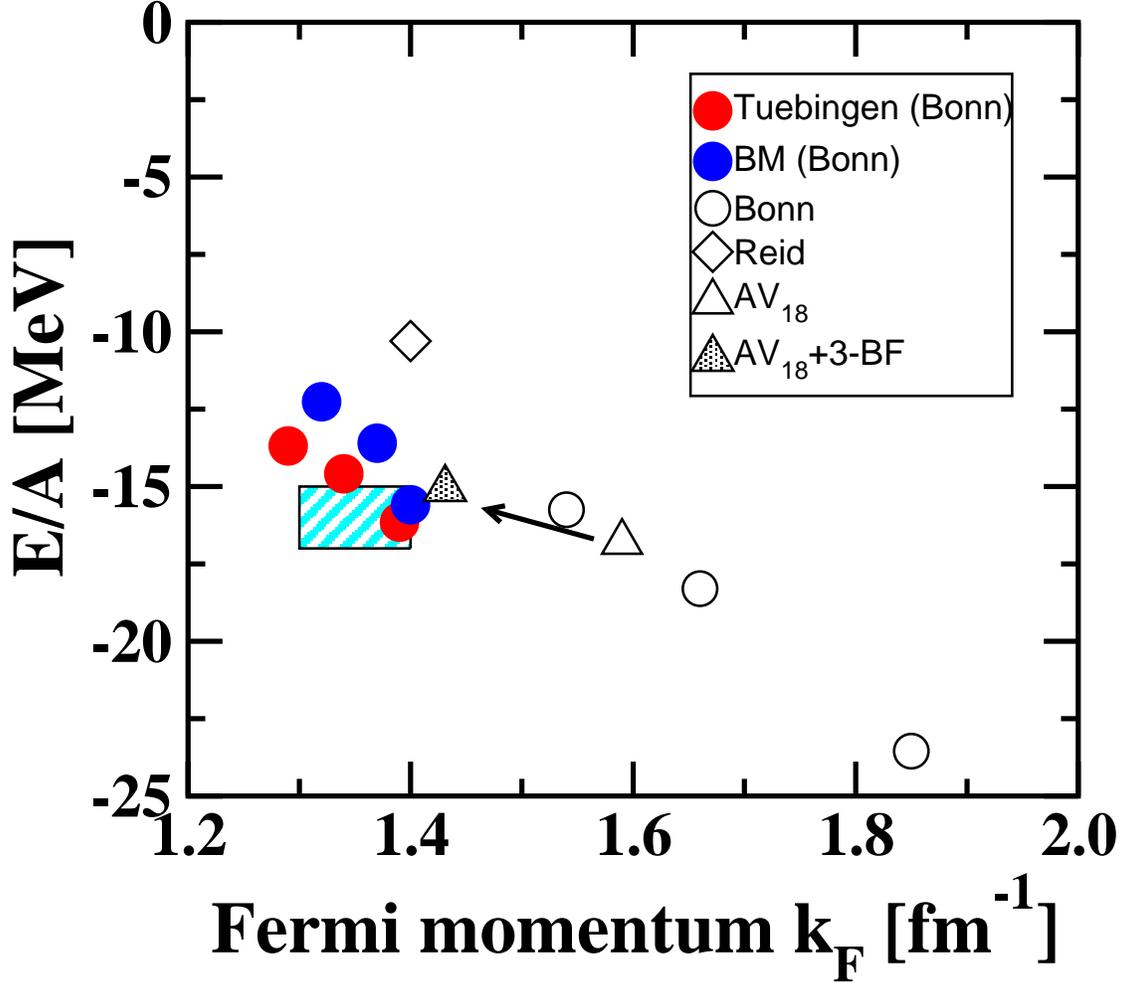}
\caption{Saturation points of relativistic DBHF calculations (full symbols) are compared to several nonrelativistic BHF calculations without three-body forces (open symbols) and to the nonrelativistic BHF calculation
including three-body forces (dotted triangular). The saturation points of the relativistic DBHF calculations are from Brockman and Machleidt (BM) and of the Tuebingen group.
\label{fig:eos_coester}}
\end{center}
\end{figure}

Furthermore, the models NL3 and DD yield rather low saturation densities around 0.15 fm$^{-3}$
whereas DDRMF saturates at a rather high density of $\rho = 0.178 \ \textrm{fm}^{-3}$.
The saturation densities of both DBHF calculations are  rather high, in particular for the calculation based on the fit method with a saturation density around 0.19 fm$^{-3}$, whereas the saturation binding energies are in good agreement with empirical values. In addition, it becomes evident from Fig.~\ref{fig:eos_DB} that, although nonrelativistic BHF calculations were able to describe the nuclear saturation mechanism qualitatively, they failed quantitatively compared to the relativistic DBHF calculations. Systematic studies for a large variety of $NN$ interactions showed that saturation points were always allocated on a
so-called Coester band in the $E/A-\rho$ plane which does not
meet the empirical region of saturation, which is demonstrated in  Fig.~\ref{fig:eos_coester}.
In particular for modern one boson exchange potentials, these nonrelativistic calculations
lead to as well too much binding as too large saturation densities. The relativistic calculations do a much better job, which can be observed  in  Fig.~\ref{fig:eos_coester} for DBHF results taken from the calculations of Brockman and Machleidt~\cite{brockmann:1990} and from the more recent calculations based on improved techniques from Ref.~\cite{gross:1999} using the Bonn potentials. The inclusion of three-body forces in the nonrelativistic BHF calculations makes the EoS stiffer, since the contributions from these forces are in total repulsive. Therefore, the inclusion of the three-body forces in these nonrelativistic BHF calculations lead to a shift of the saturation point away
from the  nonrelativistic Coester band towards its relativistic counterpart where the DBHF results are allocated. This shift of the saturation point
is demonstrated for the BHF calculations~\cite{zuo:2002} denoted by 
triangles in Fig.~\ref{fig:eos_coester}, where saturation density is shifted 
from a value of 0.265 fm$^{-3}$ to a value of 0.198 fm$^{-3}$.

\begin{figure}[!t]
\begin{center}
\includegraphics[width=0.3\textwidth] {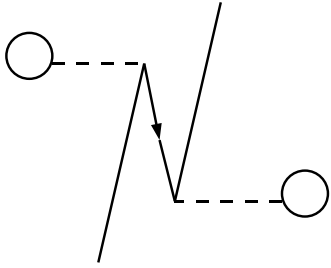}
\caption{Modification of the propagator representing the enhancement of the
small component of the nucleon Dirac spinor in the medium due to the interaction
with other nucleons.
\label{fig:zgraph}}
\end{center}
\end{figure}

One may get the impression that the three-body forces, which are required in
nonrelativistic calculations to achieve the empirical saturation point,
corresponds to a parameterisation of the relativistic features, which are
included in DBHF calculations. Indeed, if one expands the propagator of a
nucleon in the nuclear medium in terms of the propagators for the free nucleon,
the lowest order term, which represents a change of the Dirac mass or the
enhancement of the small component in the in-medium Dirac spinor, which is
described by this in medium mass, is displayed in Fig.~\ref{fig:zgraph}. This
so-called Z-graph involves two extra nucleon in addition to the propagating one and
therefore should be represented in a nonrelativistic theory, which ignores this
medium dependence of the Dirac spinor, by a three-body
force~\cite{brown:1987,bouyssy:1987}.

\begin{figure}[!t]
\begin{center}
\includegraphics[width=0.9\textwidth] {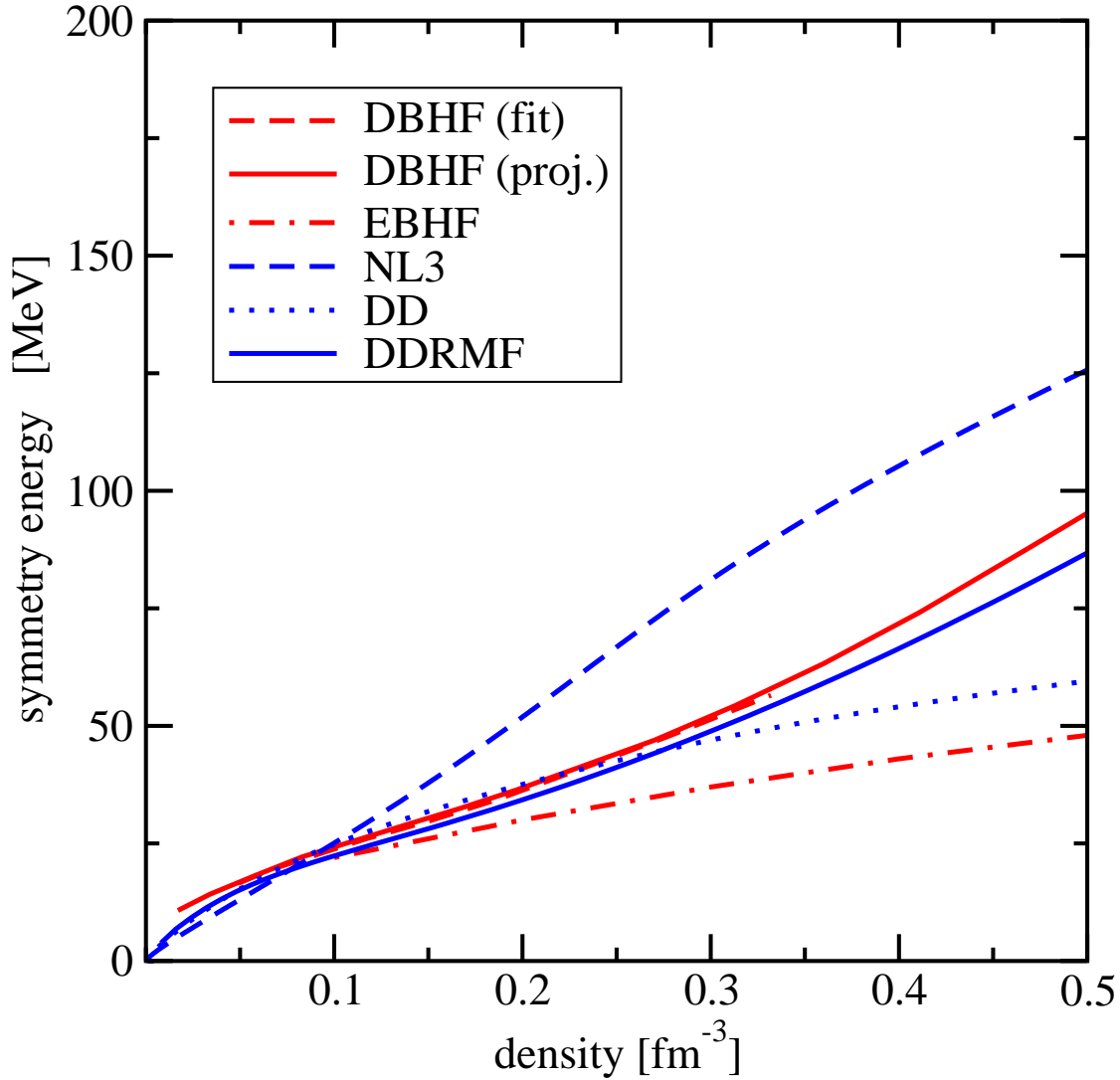}
\caption{Comparison of the symmetry energy obtained from relativistic many-body calculations.
The relativistic ab initio approaches are a DBHF using the fit method and one using projection techniques. The relativistic mean-field models presented are the nonlinear NL3 model, the DD model, and the DDRMF model. In addition, the result of a nonrelativistic extended BHF is shown.
\label{fig:sym_E}}
\end{center}
\end{figure}

The energy functional of nuclear matter can be expanded in terms of the asymmetry parameter $\beta=(\rho_n-\rho_p)/\rho$
($\rho_n$ and $\rho_p$ are the neutron and proton densities, respectively) which leads to a parabolic dependence on $\beta$
\beqa
E_b(\rho,\beta)=E(\rho)+E_{sym}(\rho) \beta^2 + {\cal O}(\beta^4).
\eeqa
The first quantity, $E(\rho)$, yields the binding in isospin symmetric nuclear matter, whereas
the quantity $E_{sym}(\rho)$ is called symmetry energy and characterizes the isospin dependence of the nuclear equation of state (EoS).
Therefore, it is defined as
\beqa
E_{\rm sym}(\rho)= \frac{1}{2} \left[
  \frac{\partial^2E_b(\rho,\beta)}{\partial \beta^2} \right]_{\beta=0}.
\eeqa
The density dependence of this symmetry energy for the models presented in Table~\ref{table:NM_Prop} is shown in
Fig.~\ref{fig:sym_E}.
It can be observed that the symmetry energies in the various models are rather
similar at a density near 0.1 fm$^{-3}$. In phenomenological models, such as RMF
models, the symmetry energy is constrained by the skin thickness of heavy nuclei
which seems to fix the symmetry energy at an average density of about 0.1
fm$^{-3}$. The results of the microscopic approaches coincide at this
density with RMF phenomenology demonstrating that the low density behavior of the
microscopic calculation is in agreement with the constraints from finite nuclei.
However, at densities above the saturation density the predictions for the 
symmetry energy are quite different for the
various many-body calculations. This high density
behavior of the symmetry energy is essential for the description of the
structure and of the stability of neutron stars. Although microscopic DBHF
calculations give a relatively soft EoS in isospin symmetric nuclear matter,
these calculations yield a rather stiff isospin dependence, respectively
symmetry energy, at high density. In general, they are significantly stiffer
than in non-relativistic BHF approaches, such as the extended BHF (EBHF)
approach~\cite{zuo:1999} using the Argonne V$_{14}$~\cite{wiringa:1984}. On the
phenomenological side, the NL3 model has the stiffest EoS in
Fig.~\ref{fig:sym_E} and the symmetry energy rises almost linearly with the
density. In contrast, the DD model exhibits a considerable flattening.
Therefore, the different phenomenological relativistic mean-field models can
yield as well a soft  as a stiff isopin dependence of the EoS.

\section{Finite Nuclei}
\label{sec:nuclei}
Many investigations have been devoted to a relativistic description of nuclei. 
Most of these studies, however, have been performed employing phenomenological
models for the $NN$ interaction. A very brief summary on such studies will be
presented in Sec.~\ref{subsec:pm}. Until now no relatvistic calculation for
finite nuclei has been presented, which is based on a realistic $NN$
interactions and evaluates both, the correlation effects and the Dirac effects
directly for the finite system under consideration (see e.g. the review by N.
Van Giai et al.~\cite{giai:2010}). In Sec.~\ref{subsec:mbm} we
are going to present the status of the approximation schemes for such DBHF
calculations in finite nuclei. Special attention will be paid to the spin orbit
splitting of the single-particle spectrum in Sec.~\ref{subsec:sos}, as the
spin-orbit term can be considered as a kind of fingerprint for relativistic
effects in nuclear systems.

\subsection{Phenomenological Methods}
\label{subsec:pm}
Very popular phenomenological models are relativistic density functional
theories, such as the RMF
theory~\cite{serot:1986,ring:1996,walecka:1974,liu:2002,lalazissis:1997,miller:1972}.
This RMF framework, or in other words, the relativistic Hartree approach with
the no-sea approximation is already discussed in Sec.~\ref{subsec:RMF}. This
approach, however, is missing the pion exchange processes due to the absence of
Fock terms. This missing pion plays an essential role in determining the isospin
dependence of the shell evolutions ~\cite{long:2008}. Therefore, considerable
effort has been devoted to the relativistic Hartree-Fock (RHF) theory to include
the Fock
terms~\cite{bouyssy:1987,brockmann:1978,horowitz:1984,bouyssy:1985,bernardos:1993,marcos:2004}.
However, it failed in a quantitative description of these nuclei for a long
time, because of its numerical complexity. After the introduction of microscopic
based density dependent coupling
functions~\cite{marcos:1989,elsenhans:1990,brockmann:1992}, phenomenological
density dependent coupling functions were recently introduced in the
relativistic Hartree (RH) theory~\cite{niksic:2002} and in the relativistic
Hartree-Fock  (DDRHF) theory~\cite{long:2006,long:2007}. It has lead to a good
quantitative description of some nuclei, as can be observed in
Table~\ref{tab:long_param}.
\begin{table}[h]
\begin{center}
\begin{tabular}{|l|ccc|ccc|}
\hline
        &           &PKA1~\cite{long:2007}&          &           & exp.~\cite{brown:1998,audi:1993,chartier:1996,fricke:1995}     &         \\
        &           &($\sigma,\omega,\pi,\rho$)&    &          &        &       \\
        &$E/A$ [MeV]&	                        &$r_c$ [fm]&$E/A$ [MeV] &        &   $r_c$ [fm] \\
\hline
\ $^{16}O$ \ & -7.94  &                           &2.80      &-7.98       &    &2.74        \\
\ $^{40}Ca$ \ & -8.54 &                           &3.53      &-8.55       &    &3.48        \\
\ $^{48}Ca$ \ &-8.67 &                           &3.49      &-8.67       &    &3.47         \\
\ $^{90}Zr$ \ &-8.71 &                           &4.28      &-8.71       &    &4.27         \\
\ $^{208}Pb$ \ & -7.87&                           &5.51      &-7.87       &    &5.50         \\
\hline
\end{tabular}
\end{center}
\caption{\label{tab:long_param}
        Results for closed shell nuclei applying the purely phenomenological
	  DDRHF parameterizations PKA1.
          Experimental values are also given.}
\end{table}
The good agreement is not so surprising, since the parameterization PKA1~\cite{long:2007} was fitted to these nuclei. It is a general feature that the parameters in these purely phenomenological theories have been adjusted to describe the saturation properties of nuclear matter or the properties of stable nuclei located in the valley of $\beta$ stability. Therefore, the predictive power of these phenomenological interactions is rather limited, in particular for exotic nuclear systems such as the neutron star crust and nuclei far away from the line of $\beta$ stability.

\subsection{Microscopic Based Methods.}
\label{subsec:mbm}

\textit{Ab initio} approaches, such as the DBHF approach, are based on high
precision realistic free space nucleon-nucleon interactions and the nuclear
many-body problem is solved without any adjustment of parameters. Therefore, 
if calculations of this kind are successful in describing properties of normal
nuclei, we have got a tool which connects properties of baryons in the vacuum
($NN$ scattering data) with nuclear systems at densities around the saturation
density of nuclear matter. Such a tool should have  a high predictive power,
when it is used in extreme cases such as nuclear matter at densities beyond the
saturation point or
in a highly isospin asymmetric nuclear environment. Although DBHF calculations
have been quite successful in describing nuclear matter, and therfore it
fulfills this requirement, full Dirac Brueckner
calculations are still too complex to allow an application to finite nuclei at
present. Therefore, different approximation schemes have been developed,
which will treat either the correlation effects or the relativistic effects in
an approximative way. These approximation schemes shall be described in this
section.

\subsubsection{Treatment of the Correlation Effects in a Local Density Approximation.}
In the first approximation scheme, the Dirac effects are treated directly for the finite nucleus, whereas the correlation effects are treated in a local density approximation. This treatment of the correlation effects can be accomplished by defining effective meson exchange interactions based upon the nuclear matter $T$ matrix. The coupling constants of such an effective interaction, which are adjusted so as to reproduce the in-medium $T$ matrix, are density dependent, since the in-medium $T$ matrix is density dependent~\cite{marcos:1989,elsenhans:1990}. In this way, a semi-phenomenological relativistic density functional can be constructed in contrast to the purely phenomenological relativistic mean-field model. These semi-phenomenological relativistic density functionals can be divided into semi-phenomenological DDRH theories and semi-phenomenological DDRHF theories.

In order to properly parameterize the DBHF self-energy components in isospin asymmetric nuclear matter, the coupling functions in the semi-phenomenological DDRH theory should be based on four different channels: scalar isoscalar, vector isoscalar, scalar isovector, and vector isovector channel. These channels correspond to phenomenological exchange bosons, i.e. the $\sigma$-, $\omega$-, $\delta$-, and $\rho$-mesons and these coupling functions  are given by
\begin{eqnarray}
\left(\frac{g_{\sigma}(\rho ,\rho_3)}{m_{\sigma}}\right)^2 = - \frac{1}{2}
\frac{\Sigsp(k_{\mathrm Fp})  + \Sigsn(k_{\mathrm Fn})}{\rho^s}, \label{eq:ss}\\
\left(\frac{g_{\omega}(\rho ,\rho_3)}{m_{\omega}}\right)^2 = - \frac{1}{2}
\frac{\Sigop(k_{\mathrm Fp})  + \Sigon(k_{\mathrm Fn})}{\rho }, \label{eq:vw} \\
\left(\frac{g_{\delta}(\rho ,\rho_3)}{m_{\delta}}\right)^2 = - \frac{1}{2}
\frac{\Sigsp(k_{\mathrm Fp})  - \Sigsn(k_{\mathrm Fn})}{\rho^s_{3}}, \label{eq:isd}\\
\left(\frac{g_{\rho}(\rho ,\rho_3)}{m_{\rho}}\right)^2 = - \frac{1}{2}
\frac{\Sigop(k_{\mathrm Fp})  - \Sigon(k_{\mathrm Fn})}{\rho_3},  \label{eq:ivr}
\end{eqnarray}
with $\rho^s =\rho^s_{p}+ \rho^s_{n}$, $\rho =\rho _p+\rho _n$,
$\rho^s_{3}=\rho^s_{p}-\rho^s_{n}$, and $\rho _3=\rho _p- \rho _n$,
where
\begin{eqnarray}
\rho^s_{i}=\frac{2}{(2\pi)^3} \int_0^{k_{Fi}} d^3{k} \frac{m^*_i}{\sqrt{{m^*_i}^2+k^2}}
\end{eqnarray}
and
\begin{eqnarray}
\rho _i=\frac{2}{(2\pi)^3} \int_0^{k_{Fi}} d^3{k} = \frac{k_{Fi}^3}{3 \pi^2}
\end{eqnarray}
are, respectively, the scalar and vector density of the particle $(i=n,p)$.

In initial DDRH calculations, such as in Refs.~\cite{fuchs:1995,brockmann:1992}, the isovector scalar $\delta$ coupling is missing. 
However, this coupling provides a mechanism to account for the differences in
the scalar self-energies and in the corresponding effective Dirac masses for neutrons and protons in isopin asymmetric nuclear matter~\cite{schiller:2001}. Therefore, it has important consequences for the dynamics of neutron-rich nuclei~\cite{dejong:1998}.

\begin{figure}[!t]
\begin{center}
\includegraphics[width=0.9\textwidth] {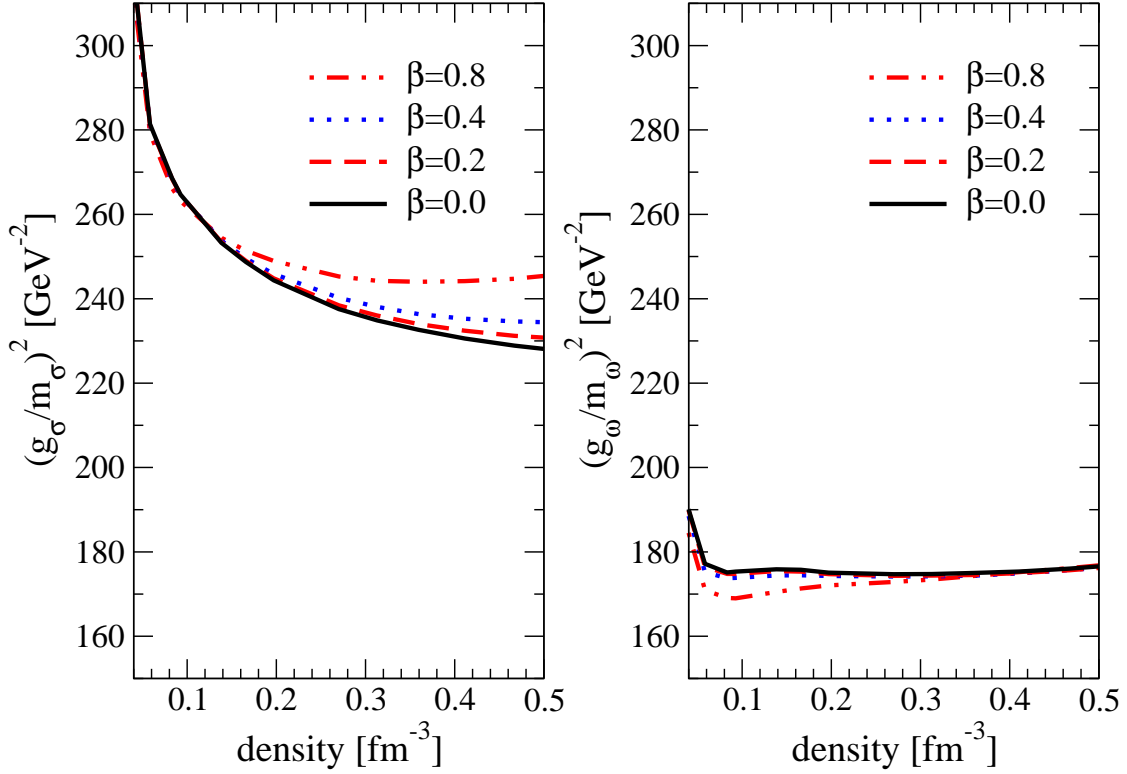}
\caption{The isoscalar effective coupling functions $g_\sigma$ and $g_\omega$ are plotted as a function of the baryon density for different values of the  asymmetry parameter $\beta$. \label{fig:sigmaomega}}
\end{center}
\end{figure}

\begin{figure}[!t]
\begin{center}
\includegraphics[width=0.9\textwidth] {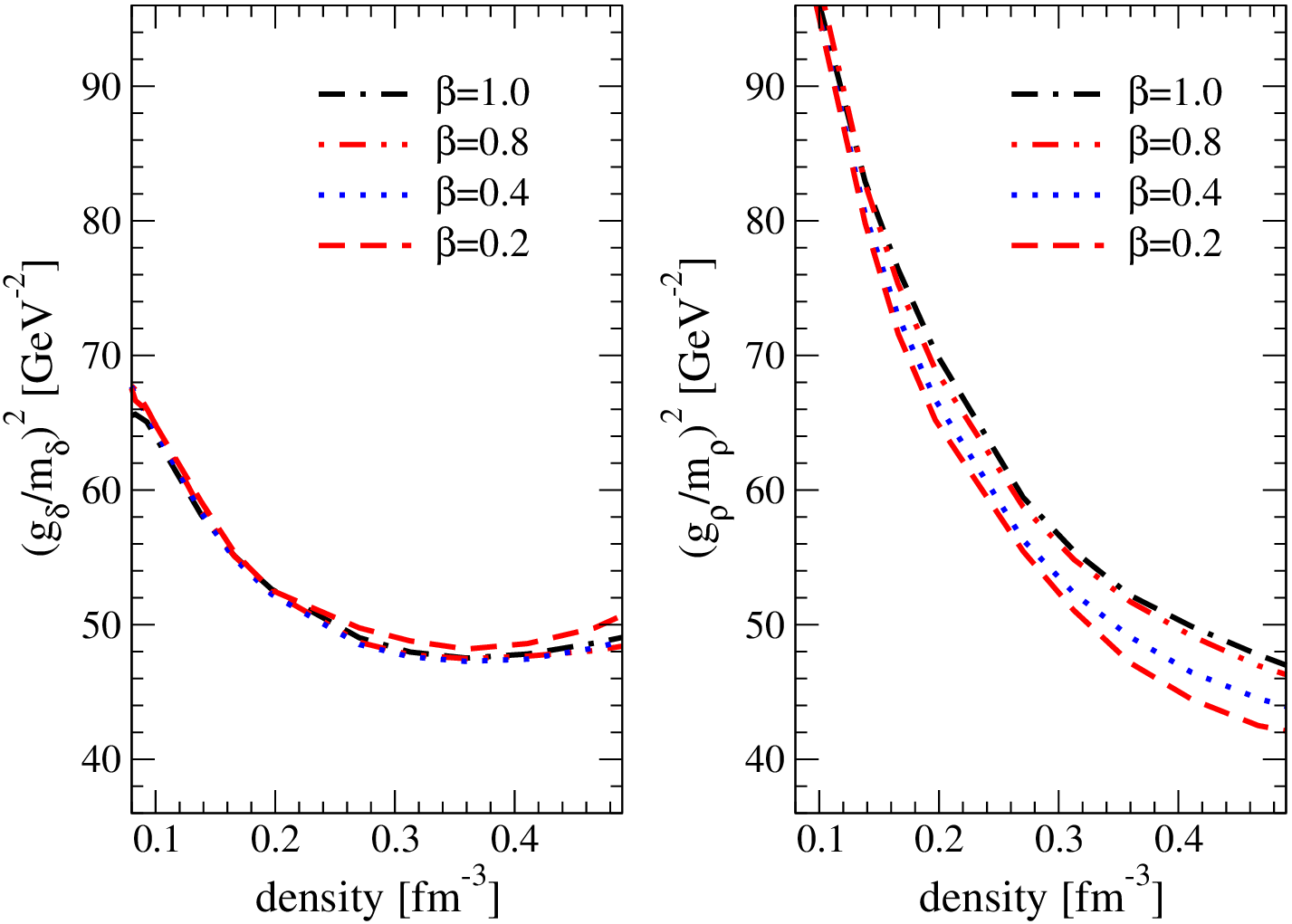}
\caption{The isovector effective coupling functions $g_\delta$ and $g_\rho$  are plottted as a function of the baryon density for different values of the  asymmetry parameter $\beta$.  \label{fig:deltarho}}
\end{center}
\end{figure}

The coupling constants of Eqs. (\ref{eq:ss})-(\ref{eq:ivr}) depend on the density $\rho$ and on the proton-neutron asymmetry density
$\rho_3$. In practice, the dependence on $\rho_3$ of the
coupling functions is in general ignored to keep the functional as simple as possible~\cite{schiller:2001,vandalen:2004b,vandalen:2007,hofmann:2001}.
The reason is the weakness of this $\rho_3$ dependence. In Fig.~\ref{fig:sigmaomega} this weak dependence on $\rho_3$ can be observed for the isoscalar coupling functions, which are obtained from the self-energy components of the DBHF approach in Ref.~\cite{vandalen:2007}. Furthermore, these coupling functions decrease with increasing density, which is attributed to the correlation effects~\cite{elsenhans:1990}.The results for the isovector coupling functions are presented in Fig.~\ref{fig:deltarho}. 


\begin{figure}[!t]
\begin{center}
\includegraphics[width=0.9\textwidth] {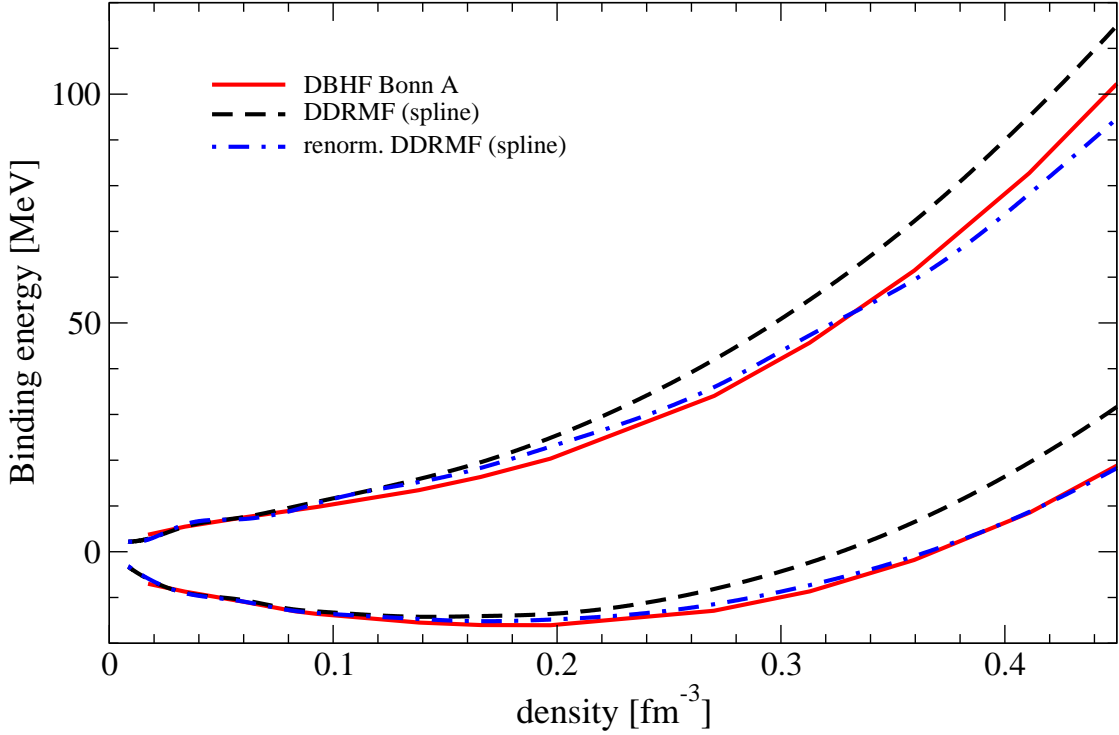}
\caption{Comparison between results of the DBHF approach (solid line), the DDRMF approach (dashed line), and the renormalized DDRMF approach (dashed-dotted line) for pure neutron (up) and symmetric (down) matter. The DDRMF approach and the renormalized DDRMF approach presented are based on the DBHF approach
using a spline interpolation.
\label{fig:EbindDBHF}}
\end{center}
\end{figure}

However, one observes clear deviations between the original DBHF EoS and the DDRH EoS based on this DBHF EoS via the parameterization (\ref{eq:ss})-(\ref{eq:ivr}). These deviations are not due to the dependence on $\rho_3$, as is shown in Fig.~\ref{fig:EbindDBHF}.  Therefore, the density dependent coupling functions should be extracted more carefully as has been done in the "naive" definition (\ref{eq:ss})-(\ref{eq:ivr}) by a kind of renormalization. This fact has already been pointed out in Ref.~\cite{vandalen:2007,hofmann:2001}. The reason of these  deviations are that two essential differences between the DBHF approach and the RMF theory concerning the structure of the self-energy, respectively the mean-field, exist.

The first difference is the appearance of a spatial contribution of the vector self-energy
$\Sigma_V$ in the DBHF theory, which is not present in the RMF model.  This self-energy
component originates from Fock exchange contributions which are not
present in the RMF theory. Therefore, this spatial $\Sigma_V$ component has to be included in a proper way
for an accurate reproduction of the properties of the original DBHF EoS. The effects of the $\Sigma_V$ component in the Dirac equation for
homogeneous nuclear matter can be absorbed into a renormalization of the scalar
and time-like vector component of the self-energy~\cite{vandalen:2007,gogelein:2008}. Firstly, the $\Sigma_V$ component can be absorbed into the effective mass according to Eq.~(\ref{subsec:SM;eq:redquantity}) and this reduced effective mass has to be identified with the RMF effective mass, i.e.
\begin{eqnarray}
\tilde{m}^{*}_i=\frac{M + \Sigsi(k_{Fi})}{1+\Sigvi(k_{Fi})}=M+\Sigma^{DDRMF}_{{\mathrm s},i}.
\label{eq:meffDBHF}
\end{eqnarray}
This leads to the renormalized scalar self-energy component
\begin{eqnarray}
\Sigma^{DDRMF}_{{\mathrm s},i}=\frac{\Sigsi(k_{Fi})-M \Sigvi(k_{Fi})}{1+\Sigvi(k_{Fi})}.
\end{eqnarray}
In a corresponding way, however, using the energy density instead of the effective mass, the following expression for the normalized vector
self-energy component is obtained
\beqa
\SigoDDRMFi=\Sigoi(k_{Fi})-\frac{\Sigvi(k_{Fi}) [3 E_{Fi} \rho_i +  \tilde{m}^*_i \rho_{s,i}]}{4 \rho_i}.
\eeqa
These renormalized self-energy components are now inserted into Eqs.~(\ref{eq:ss})-(\ref{eq:ivr}) to obtain the renormalized density dependent coupling functions. This renormalized DDRMF theory yields nuclear matter properties, which are closer to those of the original EoS~\cite{vandalen:2007}  as can be observed in Fig.~\ref{fig:EbindDBHF}. Such a theory also provides good results for nuclei, as can be seen in Table~\ref{tab:FN_vD_param} for the renormalized DDRMF theory in Ref.~\cite{gogelein:2008}.

\begin{table}[h]
\begin{center}
\begin{tabular}{|l|ccc|}
\hline
        &           &DDRMF~\cite{gogelein:2008}&          \\
        &$E/A$ [MeV]&	                        &$r_c$ [fm] \\
\hline
\ $^{16}O$ \ & -8.35  &                           &2.67      \\
\ $^{40}Ca$ \ & -8.73 &                           &3.35      \\
\ $^{48}Ca$ \ &-8.73 &                           &3.39      \\
\ $^{90}Zr$ \ &-8.74 &                           &4.12      \\
\ $^{208}Pb$ \ & -7.87&                           &5.30     \\
\hline
\end{tabular}
\end{center}
\caption{\label{tab:FN_vD_param}
          Results for closed shell nuclei applying the
	  DDRMF parameterization from the DBHF results.}
\end{table}

The second difference is that the DBHF self-energy terms explicitly depend on the momentum of the particle, a
feature which is absent in RMF. This momentum dependence reflects the non-locality of the DBHF
self-energy terms, which originates not only from the Fock exchange terms but also from
non-localities in the underlying $NN$ interaction. In the work of Ref.~\cite{hofmann:2001}, a method was introduced to
compensate for this momentum dependence. Therefore, the full DBHF self-energies are expanded around the
Fermi momentum,
\begin{equation}
\Sigma(k,\kf)=\Sigma(\kf,\kf)+(k^2-\kf^2) \frac{\Sigma(k,\kf)}{\partial k^2} \biggr{|}_{k=\kf} + \bigcirc(k^4).
\label{eq:expansigma}
\end{equation}
The first term in Eq.~(\ref{eq:expansigma}) can be regarded as the Hartree self-energy, whereas the second term is a measure of
the momentum dependence around the Fermi surface. In the next step, one should identify the energy density of DBHF EoS with that of the DDRH theory.
Furthermore, the momentum will be integrated out by averaging over the Fermi sphere,
\begin{eqnarray}
\rho \SigoDDRH(\kf)& = & \frac{4}{(2 \pi)^3} \int_{|k| \leq \kf} d^3k \Sigo(k,\kf) \nn \\
                       & = & \rho \Sigo(k,\kf) \biggl{[}1-\frac{2}{3} \kf^2 \frac{\Sigo'(\kf)}{\Sigo(\kf)} \biggr{]}.
\end{eqnarray}
Instead of extracting
\begin{equation}
\Sigo'(\kf)=\frac{\Sigma(k,\kf)}{\partial k^2} \biggr{|}_{k=\kf}
\end{equation}
directly from DBHF calculations, $\Sigo'(\kf)$ is calculated numerically by adjusting the DDRH binding energy to that of the DBHF EoS in Ref.~\cite{hofmann:2001}. Therefore, this correction of the self-energy can be translated into a modification of the density dependent coupling functions,
\begin{equation}
\tilde{g}^2(\kf)=g^2(\kf) \biggl[ 1-\frac{2}{3} \kf^2 \frac{\Sigo'(\kf)}{\Sigo(\kf)} \biggr].
\end{equation}
Making the assumption that the ratio $\frac{\Sigo'(\kf)}{\Sigo(\kf)}$ only depends weakly on the Fermi momentum,
one can introduce the following momentum corrected coupling functions,
\begin{equation}
\tilde{g}_{\alpha}(\kf)=g_{\alpha}(\kf) \sqrt{1-\zeta_{\alpha} \kf^2},
\end{equation}
where the constants $\zeta_{\alpha}$ are determined by fitting to the DBHF EoS. This momentum dependence correction improves
the calculated results as can be observed from Table~\ref{tab:FN_Hoff_param} for some closed shell nuclei.
\begin{table}[h]
\begin{center}
\begin{tabular}{|l|ccc|ccc|}
\hline
        &           &DDRH~\cite{hofmann:2001}&          &           &cor. DDRH~\cite{hofmann:2001}&           \\
        &$E/A$ [MeV]&    &$r_c$ [fm] &$E/A$ [MeV]&         &$r_c$ [fm] \\
\hline
\ $^{16}O$ \ &-8.58       &    &2.75      &-8.30      &         &2.79  \\
\ $^{40}Ca$ \ &-9.02       &    &3.46      &-8.69      &         &3.50  \\
\ $^{48}Ca$ \ &-8.96       &    &3.49      &-8.63      &         &3.53  \\
\ $^{90}Zr$ \ &-8.98       &    &4.26      &-8.63      &         &4.31  \\
\ $^{208}Pb$ \ &-8.17       &    &5.53      &-7.82      &         &5.60  \\
\hline
\end{tabular}
\end{center}
\caption{\label{tab:FN_Hoff_param}
          Results for closed shell nuclei applying the
	  DDRH parameterizations without and with a correction for the momentum dependence.
           These parameterizations are based on  DBHF calculations using Bonn A.}
\end{table}

These essential differences between  the DBHF approach and the RMF theory concerning the structure of the self-energy mainly originate from the omission of the Fock exchange terms. An improvement, therefore, was the density dependent relativistic Hartree-Fock  (DDRHF) theory with its inclusion of the Fock terms. Since Fock terms are present, one obtains a spatial contribution of the vector self-energy $\Sigma_V$  in the DDRHF theory and momentum dependent
self-energy components as in the DBHF approach. Up to now the relativistic DBHF in-medium $T$ matrix is presented
by an effective interaction Lagrangian in the DDRHF theory, which only contains the $\sigma$-,$\omega$-,$\pi$-, and the $\rho$-meson~\cite{marcos:1989}.
The corresponding self-energy components in this DDRHF theory are
\beqa
\Sigsi (\veck) & = &  \frac{1}{4} \sum_{j}
\int_0^{\kfj} \frac{d^3\vecq}{(2 \pi)^3}  \frac{m^*_{j}}{E^*_{q,{j}}}
[-4 \frac{g_{\sigma}^2}{m_{\sigma}} + \delta_{ij} [ D_{\rm \sigma}(k,q) - 4 D_{\rm \omega}(k,q) ] \nn
\\ & &  - (2-\delta_{ij})[\frac{m_{j}^{*2} +
m_{i}^{*2} -  + 2 k^{* \mu} q^*_{\mu}}{(m_{i}^*+m_{j}^*)^2} D_{\rm \pi}(k,q) + 4 D_{\rm \rho}(k,q)]], \\ & & \nn \\
%
\Sigoi (\veck) & = & \frac{1}{4} \sum_{j}
\int_0^{\kfj} \frac{d^3\vecq}{(2 \pi)^3}
[- 4 \frac{g_{\omega}^2}{m_{\omega}} - 4 \frac{g_{\rho}^2}{m_{\rho}} \tau_3 - \delta_{ij} [ D_{\rm \sigma}(k,q) + 2 D_{\rm \omega}(k,q) ] \nn
\\ & & - (2-\delta_{ij}) [2 D_{\rm \rho}(k,q) -  \frac{2 E^*_{k,i} (m_{j}^{*2}-k^{* \mu}
q^*_{\mu}) - E^*_{q,j} (m^{*2}_{j} - m^{*2}_{i})}{E^*_{q,j} (m^*_{i} +
    m^*_{j})^2} \\ & &  D_{\rm \pi}(k,q)]], \nn  \\ & & \nn \\
%
\Sigvi (\veck) & = & \frac{1}{4} \sum_{j}
\int_0^{\kfj} \frac{d^3\vecq}{(2 \pi)^3} \frac{\vecq \cdot \veck}{|\veck|^2 E^*_{q,j}}
[- 4 \frac{g_{\omega}^2}{m_{\omega}} - 4 \frac{g_{\rho}^2}{m_{\rho}} \tau_3 \nn \\ & &
- \delta_{ij} [ D_{\rm \sigma}(k,q) + 2 D_{\rm \omega}(k,q) ]  - (2-\delta_{ij}) [2 D_{\rm \rho}(k,q) \nn
\\ & &
- \frac{2 k^*_z (m^{*2}_j - k^{* \mu}
  q^*_{\mu} ) - q_z (m^{*2}_j - m^{*2}_i)}{q_z (m^*_i+m^*_j)^2}
D_{\rm \pi}(k,q)]],
\eeqa
with the meson propagator,
\begin{equation}
D_k(k,q)=\frac{g_k^2}{-(k^{*\mu}-q^{*\mu})(k^*_{\mu}-q^*_{\mu})+m_k^2}.
\end{equation}
A typical feature of this theory compared to Hartree versions is the presence of a pion exchange potential. This pion exchange potential does not appear in Hartree theories, since this potential only contributes to the Fock terms. Nevertheless, these DDRHF models based on microscopic approaches are not so successful yet as DDRH models, as can be concluded from Table~\ref{tab:DDRHF_param}.
\begin{table}[b]
\begin{center}
\begin{tabular}{|l|ccc|ccc|}
\hline
        &           &HF2~\cite{fritz:1994}&          &           &RDHF3B~\cite{shi:1995}&           \\
        &           &$(\sigma,\omega,\pi)$&          &           &$(\sigma,\omega,\pi,\rho)$&           \\
        &$E/A$ [MeV]&    &$r_c$ [fm]&$E/A$ [MeV]&         &$r_c$ [fm] \\
\hline
\ $^{16}O$ \ &-7.73       &    &2.48       &-7.57      &         &2.59  \\
\ $^{40}Ca$ \ &-8.09       &    &3.14      &-7.94      &         &3.26  \\
\ $^{48}Ca$ \ &-7.90       &    &3.16      &-7.70      &         &3.27  \\
\ $^{90}Zr$ \ &-       &    &-             &-7.78      &         &3.99  \\
\ $^{208}Pb$ \ &-       &    &-            &-6.61      &         &5.17  \\
\hline
\end{tabular}
\end{center}
\caption{\label{tab:DDRHF_param}
          Results for closed shell nuclei applying the
	  DDRHF parameterizations from DBHF results.}
\end{table}
A possible reason is that only the isoscalar coupling functions are density dependent, whereas the coupling functions of the $\pi$-meson and the $\rho$-meson remain density independent in these DDRHF theories~\cite{fritz:1994,shi:1995}. Another observation is that the $\delta$-meson is not present in these DDRHF theories. Therefore, these theories can be further improved by including the $\delta$-meson and extent the density dependence to all
coupling functions, which has not been done until now for DDRHF theories based on microscopic calculations such as the DBHF approach.

\subsubsection{Treatment of the Dirac Effects in a Kind of  Local Density Approximation.}
In this scheme~\cite{fritz:1993,muether:1988,muether:1990}, the information
about the structure of the Dirac spinors is taken from the investigations of
nuclear matter. This means that the Dirac effects 
are taken into account via a kind of local density
approximation (LDA). Therefore, this approximation scheme will be called Dirac
LDA. On the other hand, the correlations effects are taken into account by
solving the Bethe-Goldstone equation directly for the finite nucleus under
consideration. This means that the
self-consistency requirements of a conventional BHF calculation for finite
nuclei are fulfilled, whereas relativistic effects
are taken into account by evaluating the potential matrix elements in terms of
in-medium Dirac spinors. 

\begin{figure}[!t]
\begin{center}
\includegraphics[width=0.99\textwidth]{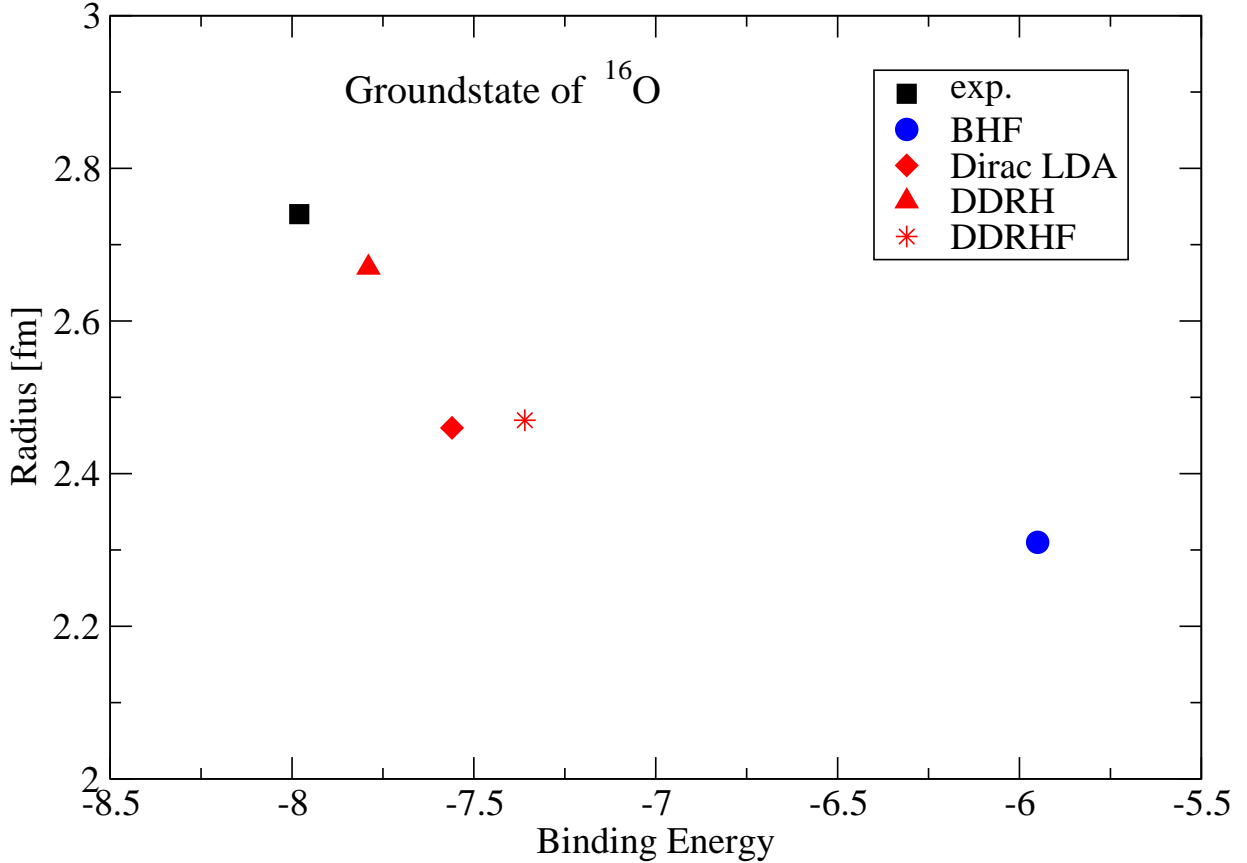}
\caption{The binding energy per nucleon and radius of the charge distribution obtained from a nonrelativistic BHF approach and some approaches based on the DBHF results are compared to the experimental value. These approaches based on DBHF results are the Dirac LDA, the density dependent relativistic Hartree (DDRH) theory, and the density dependent relativistic Hartree-Fock (DDRHF) theory.
\label{fig:bindingO16}}
\end{center}
\end{figure}

In this  Dirac LDA  scheme~\cite{fritz:1993}, the
calculated binding energy and the radius of the charge distribution is increased
compared to the results of nonrelativistic BHF calculations~\cite{fritz:1993},
as can be observed in Fig.~\ref{fig:bindingO16} in case of the isospin symmetric
$^{16}$O nucleus. 
Therefore, the results of these Dirac LDA  calculations are generally in better
agreement with experiment than the ones of the nonrelativistic BHF calculations,
in which the medium dependence of the Dirac spinors is ignored. In addition one
can see from Fig.~\ref{fig:bindingO16} that the results of this Dirac LDA
approach are very close to those obtained within the DDRHF calculation. The same
feature has also been observed for other isospin symmetric nuclei. This can
be considered as a mutual support of these two approximation schemes indicating
that LDA is a good approximation for treating correlation effects as well as
Dirac effects. The result of a complete DBHF calculation should be in the
vicinity of the Dirac LDA and DDRHF results.

On the other hand it
can be observed in Fig.~\ref{fig:bindingO16} that the result of the
semi-phenomenological DDRH calculation of Ref.~\cite{fritz:1993} is in much
better agreement with the experimental value than that of the Dirac LDA
calculation. However, this good agreement with the experimental value seems just
fortuitous, since it is lost when Fock terms are included. 

It should be recalled that all these calculations discussed in this subsection
ignore rearrangement terms, although e.g. the density dependence of the
effective coupling constants in the DDRHF approximation scheme would call for
such rearrangement terms. They have been dropped because the aim of
this study has been to develop a reliable approximation scheme for DBHF in
finite nuclei. The inclusion of rearrangement terms or the corresponding
extended microscopic definition of the nucleon self-energy would be one possible
step for future investigations to improve the results of relativistic
calculations beyond the DBHF approach.

\subsection{Spin Orbit Splitting}
\label{subsec:sos}

In atomic physics, the spin orbit splitting of the single-particle spectrum is a
genuine relativistic effect. Therefore it is quite natural to consider the spin
orbit splitting in nuclear system as a possiblle candidate to probe the
relevance of relativistic features for the nuclear many-body system. Therefore 
this section is devoted to the discussion of the spin orbit term. 
Reducing the Dirac equation for a nucleon in a RMF theory to a non-relativistic
Schr\"odinger equation one obtains a spin orbit term of the form
\begin{equation}
U_{ls}(r)=-\frac{1}{2m} \frac{D'(r)}{D(r)} \frac{\mathbf{s \cdot l}}{r},
\label{eq:uls}
\end{equation}
with
\begin{equation}
D(r)=m + \Sigma_{\mathrm s}(r)+ E  + \Sigma_{\mathrm o}(r).
\label{eq:dr}
\end{equation}
From the term $\mathbf{s \cdot l}$ in Eq.~(\ref{eq:uls}), it is expected that
the spin orbit splitting tends to become larger with increasing orbital angular
momentum. However, inspecting the experimental data for the example of the nucleus
$^{16}$O, one finds that the splitting of the $p$ orbits is larger than the
splitting of the
$d$ orbits as can be observed from Table~\ref{tab:spinorbit}.
\begin{table}[h]
\begin{center}
\begin{tabular}{|l|c|c|c|c|}
\hline
        &BHF & BHF  + 2h1p  + 1h2p       & BHF + 2h1p + 1h2p    & experiment\\
        &    &                           & + Dirac effects      &            \\
\hline
\ $\Delta\epsilon_p$ \ &3.95       & 4.11   & 6.19  &  6.18  \\
\ $\Delta\epsilon_d$ \ & 5.80      & 5.50   & 5.50  &  5.09  \\
\hline
\end{tabular}
\end{center}
\caption{\label{tab:spinorbit}
 Spin orbit splitting in $^{16}$ O. The spin orbit splitting for the $d$ shell and the $p$ shell in the various approximation schemes are given in MeV.}
\end{table}

Nonrelativistic BHF calculations can not reproduce these experimental
results for the spin orbit splitting, i.e. it predicts a larger spin orbit
splitting for the $d$ shell than for the $p$ shell~\cite{zamick:1992}. 
It is well known that contributions beyond the BHF definition of the
single-particle potential (rearrangement terms) can modify the spin orbit
splitting to quite some extent. Therefore the BHF
contribution has been supplemented with the 2-hole 1-particle (2h1p) term and with
the 2-particle 1-hole (2p1h) term. The effect of the 2h1p contribution is
repulsive and essentially cancels the effects of the 2p1h term, which is
attractive. Therefore, the inclusion of these terms do not change the spin orbit
picture in a significant way~\cite{zamick:1992}.

The Dirac LDA method is used to investigate the influence of the relativistic
effects on the spin orbit splitting. The inclusion of these relativistic effects
lead to a drastic enhancement of the spin orbit splitting for the $p$ shell,
whereas the Dirac effects  do not really enhance the spin-orbit splitting of the
$d$ shell~\cite{zamick:1992}, as can be deduced from Table~\ref{tab:spinorbit}.

A simple explanation for these findings can be provided: The spin orbit
splitting is to the first order proportional to the inverse of
m$^*$~\cite{brockmann:1977,brown:1990}, since the value of D(r) in
Eq.~(\ref{eq:dr}) can be approximated by 2m$^*$(r). In Fig.~\ref{fig:Diracsplit}
we have displayed the density dependence of the Dirac mass. So its evident that
this Dirac mass, m$^*$, is much smaller
in the dense interior of the nucleus than at the lower densities at the surface.
Therefore, the spin orbit splitting of the $p$ shell is much stronger enhanced
than the one of the $d$ shell, the latter having the largest amplitudes at the
low density region near the surface. Thus, this calculation including the
relativistic effects predicts that the spin orbit splitting in $^{16}$O for the
$p$ shell is larger than the one for the $d$ shell, which is in agreement with
experimental measurements.

Of course there also exist other approaches which reproduce the
spin-orbit effects. As an example we mention the non-relativistic
many-body calculation on $^{16}O$ by Pieper and 
Pandharipande~\cite{vjpan}. Also they can reproduce
the spin-orbit splitting of the $p$-shell, if the effects
of a 3-body force are taken into account. Again we have this feature that
effects obtained within the Dirac phenomenology are described within a
non-relativistic approach using a 3-nucleon force. It is not clear,
however, whether this 3-body force will also lead to a proper result
for the splitting in the $d$-shell.

\section{Saturation with Low-Momentum Interactions and Relativistic Effects.}

As we have already discussed in the preceeding parts it is still a challenge for
theoretical nuclear physics to predict the properties of finite nuclei from the
knowledge of a bare NN interaction, which is fitted to describe the two-nucleon
data. The DBHF approach, which yields a rather good description of the
saturation properties of symmetric nuclear matter without the necessity to
introduce a three-nucleon force, has not yet been extended to
be applicable also to finite nuclei. The reason for this missing extension is
the fact that it is not easy to evaluate the correlation effects as well as the
relativistic effects consistently for finite nuclei (see Sec.~\ref{subsec:mbm}
and Ref.~\cite{giai:2010}). 

A possible way out of this problem is the use of realistic interactions, which
are renormalised to be used in a Hilbert space with low relative momenta between
the interacting nucleons. In this way all short-range components of the NN
intteraction are integrated out and there is no need for a non-perturbative
treatment of NN correlations. Such restrictions of the nuclear structure
calculation to the low-momentum
components~\cite{Bogner01,Bozek06,bogner:2005,bogner:2007} have become very
popular during the last year within the nonrelativistic framework. 
In this section we would like to present first studies, which employ these
technique also for the relativistic approach.  

\subsection{The Low-Momentum Interaction.}
\label{subsec:LMI}

The idea of restricting the nuclear structure calculation to the low-momentum components is very reasonable.
The long-range or low-momentum part of the
$NN$ interaction is fairly well described in terms of meson-exchange, whereas
the quark degrees of freedom are more important in describing the
short-rang or high-momentum components of the $NN$ interaction.
Therefore, it is
quite attractive to disentangle these low-momentum and high-momentum components
from each other to construct a low-momentum potential. This means that a model space should be defined, which
accounts for the low-momentum degrees of freedom and renormalizes the
effective Hamiltonian for this low-momentum regime to account for the effects of
the high-momentum parts. One of the most used
methods to disentangle these parts is the unitary-model-operator approach (UMOA)~\cite{Suzuki82}.

Therefore, one has to define two projection operators.
As in all well-known model space techniques, the operator $P$ projects onto the model space of two-nucleon wave functions with
relative momenta $k$ smaller than a chosen cut-off $\Lambda$,
whereas the operator $Q$ projects onto the complement of this subspace, the high-momentum subspace.
In addition, these operators P and Q satisfy the usual
relations like $P+Q=1$, $P^2=P$, $Q^2=Q$, and $PQ=0=QP$.
In the UMOA approach, the aim is to define a unitary transformation $U$ in
such a way that the transformed Hamiltonian does not couple $P$ and $Q$, i.e.,
\begin{equation}
 QU^{-1}HUP = 0.
\label{eq:umoa}
\end{equation}
The technique to determine this unitary transformation has
nicely been described in Ref~\cite{Bozek06,fuji:2004}.
With the help of this unitary transformation an effective Hamiltonian $H_{eff} = h_0 + V_{eff}$ can be obtained.
It contains the kinetic energy $h_0$ and an effective interaction $V_{eff}$
given by
\begin{equation}
V_{eff} = V_{lowk}= U^{-1}(h_0 + V) U - h_0.
\label{eq:Vlowk1}
\end{equation}

The eigenvalues, which are obtained in the diagonalization of  this
effective Hamiltonian in the $P$-space, are identical to those, which are
obtained by diagonalizing the original Hamiltonian $H=h_0+V$ in the
complete space. This means that solving the Lipmann-Schwinger equation for NN scattering using this $V_{eff}$ with a cut-off
$\Lambda$ yields the same phase shifts as obtained for the original realistic interaction
$V$ without a cut-off. Therefore, this $V_{eff}$, which from now on will be called $V_{lowk}$,
can also be regarded as a realistic interaction like e.g. the CD Bonn or Argonne V18 interactions,
since it reproduces the $NN$ scattering phase shifts.

An interesting feature is that the resulting $V_{lowk}$ is found to be essentially
independent of the underlying realistic interaction $V$, when $V$ is fitted to the
experimental phase shifts and the cut-off $\Lambda$ is chosen around
$\Lambda$ = 2 fm$^{-1}$. Hence, one is able to obtain a low-momentum potential
$V_{lowk}$, which in a model independent manner describes the low-momentum
component of realistic $NN$ interactions in a more or less unique way.

Furthermore, the $V_{lowk}$ does not induce any short-range correlations into the nuclear wave function,
because the high-momentum components have been integrated
out by means of the unitary transformation of Eq.(\ref{eq:umoa}).
Therefore, mean-field calculations using $V_{lowk}$ already
lead to reasonable results and corrections of many-body theories beyond mean-field are weak~\cite{Bozek06}.

\subsection{Saturation Behavior of $V_{lowk}$}

\begin{figure}[!t]
\begin{center}
\includegraphics[width=0.8\textwidth] {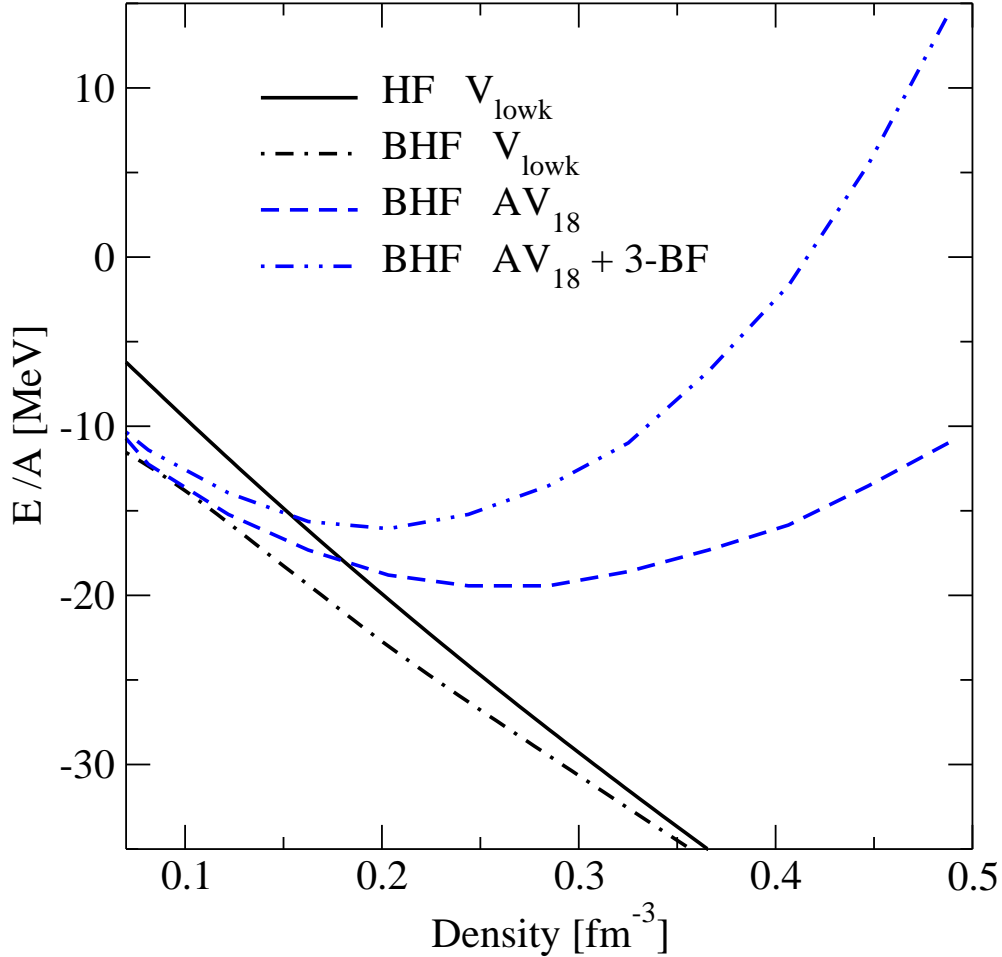}
\caption{Binding energy per nucleon of symmetric nuclear matter of a HF
calculation (solid line) and of a BHF calculation (dashed dotted line) employing a $V_{lowk}$ interaction are plotted.
In addition, the binding energy per nucleon of a nonrelativistic BHF
with three-body force calculations (dashed double dotted line)
and without three-body force (dashed line)  is presented using the Argonne $V_{18}$ potential.
\label{fig:EbVlowk}.}
\end{center}
\end{figure}

An important feature of an EoS is the saturation behavior.
However, the saturation behavior of $V_{lowk}$ is problematic, since in HF calculations of nuclear matter one obtains a binding energy
per nucleon increasing with density in a monotonic
way~\cite{vandalen:2009a,vandalen:2009b}, as can be seen in Fig.~\ref{fig:EbVlowk}.
Hence, HF calculations employing $V_{lowk}$ do not exhibit a minimum in the
energy as a function of the density and the emergence of a saturation point is
prevented in isospin symmetric nuclear
matter~\cite{Bozek06,Kuckei03,vandalen:2009a,gogelein:2009,vandalen:2009b}. This
absence of saturation also leads to problems in the description of finite
nuclei. In fact, Hartree-Fock calculations of finite nuclei using $V_{lowk}$
yield higly compressed nuclei, if the variational space is not restricted~\cite{vandalen:2009b}.

This absence of saturation is one of
the major problems of calculations employing $V_{lowk}$. Even the inclusion of correlations beyond the HF
approximation, e.g. by means of the BHF approximation~\cite{vandalen:2009a}, can not solve this
problem as can be observed from Fig.~\ref{fig:EbVlowk}.
Employing a conventional model for a realistic
interaction, like e.g. the Argonne V18, which is not
reduced to its low-momentum components, one obtains a saturation point in
calculations which account for correlations beyond the mean-field approximation~\cite{zuo:2002}  (see Fig.~\ref{fig:EbVlowk}).
Therefore, one may argue that the lack of
short-range correlation effects, which have been integrated
out by means of the unitary transformation of Eq. (\ref{eq:umoa}), prevents the emergence of a saturation point in calculations of symmetric
nuclear matter~\cite{Kuckei03,bogner:2005}. This means that the $V_{lowk}$ approach can not reproduce the
saturation of nuclear matter as it misses the quenching of short-range
correlations in the nuclear medium. This quenching of short-range
correlations is included in sophisticated
calculations employing one of the conventional models for a realistic $NN$ interaction.

Although calculations, like Brueckner-Hartree-Fock (BHF), for these conventional realistic interactions lead to a saturation point,
they are not able to reproduce the
experimental data as already mentioned in Sec.~\ref{sec:sat}. In order to obtain a saturation point closer to the empirical data one has to include a
three-body force~\cite{lejeune:2000} or include relativistic effects, e.g. within
a Dirac-Brueckner-Hartree-Fock (DBHF) approach~\cite{vandalen:2004b,vandalen:2007,muether:1990,brockmann:1984}.
Since it can be expected from these findings that these measures will also improve the saturation behavior of calculations employing  $V_{lowk}$,
this issue has been investigated. In Ref.~\cite{bogner:2005}, it is shown that the addition of three-body forces can lead to saturation in nuclear matter. The inclusion of the relativistic effects will now  be elaborately discussed in Sec.~\ref{subsec:releffect}.

\subsection{Relativistic Effects}
\label{subsec:releffect}

Relativistic effects lead to a successful microscopic description of the
saturation properties of nuclear matter in the DBHF approach. The main reason
for this success can be attributed to the fact that the matrix elements of the
bare nucleon-nucleon interaction become density dependent, since the spinor
contains the reduced effective mass~\cite{anastasio:1983}. This density
dependence is absent in nonrelativistic Hartree-Fock or Brueckner calculations.

\begin{figure}[!t]
\begin{center}
\includegraphics[width=0.8\textwidth] {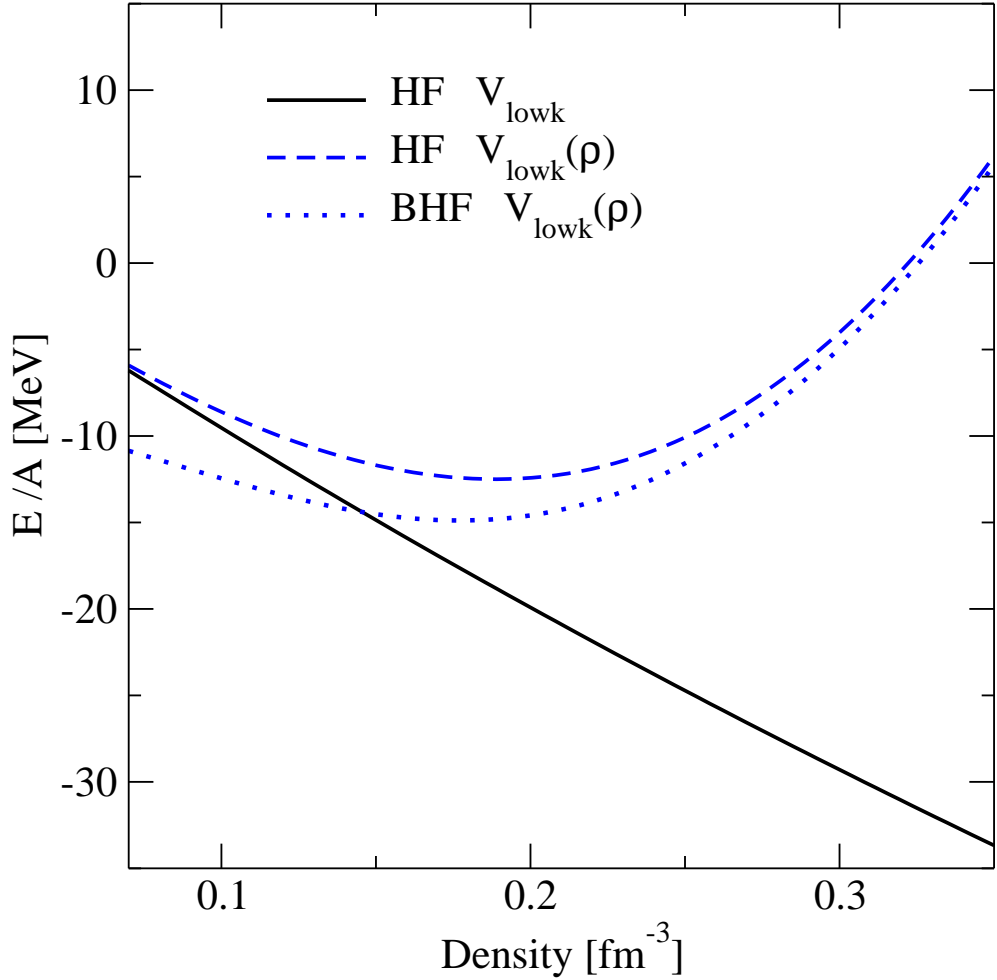}
\caption{Binding energy per nucleon of isospin symmetric nuclear matter of a HF
calculation (dashed line) and of a BHF calculation (dotted line) employing a density dependent $V_{lowk}$ interaction are plotted. In addition, a HF calculation using a standard $V_{lowk}$ interaction (solid line) is added to the figure.
\label{fig:EbDDVlowk}}
\end{center}
\end{figure}

These relativistic effects are introduced in  $V_{lowk}$ leading to a density dependent $V_{lowk}(\rho)$~\cite{vandalen:2009a}.
This is achieved  by calculating for each density the underlying realistic interaction in
terms of these dressed Dirac spinors. The density dependent $V_{lowk}(\rho)$ can then be obtained using the standard techniques given in Sec.~\ref{subsec:LMI}. Results for the energy per nucleon of isospin symmetric nuclear matter as a function of
the density $\rho$ obtained from HF and BHF calculations employing density dependent $V_{lowk}(\rho)$~\cite{vandalen:2009a}  are displayed in
Fig.~\ref{fig:EbDDVlowk}.
The density dependent $V_{lowk}(\rho)$ presented in this figure is based on the matrix elements of the Bonn A potential using
the Dirac spinors, which are appropriate for this density. Therefore, the medium properties of the nucleons, which are used to dress these Dirac spinors, are taken from the EoS presented in Refs.~\cite{vandalen:2004b,vandalen:2007}. Employing this density dependent $V_{lowk}(\rho)$ interaction with a cut-off $\Lambda$ of  2 fm$^{-1}$, one already obtains a saturation point in the HF calculation~\cite{vandalen:2009a}, as can be observed from Fig.~\ref{fig:EbDDVlowk}.
The reason is that the attractive contributions of the scalar meson $\sigma$-meson are reduced, because the small component in the Dirac spinor
is enlarged due to the fact that the effective Dirac mass is smaller than the bare mass. Because this effect grows with increasing density, the HF calculation employing the density dependent $V_{lowk}(\rho)$ interaction already leads to a minimum in the energy as a function of the density.
Although this HF calculation yields a saturation density in the neighbourhood of the empirical region, it still provides too little binding, i.e. about
-12 MeV per nucleon.

This HF result can be improved by taking into account the effects
of $NN$ correlations within the BHF approximation, which yields about 2 MeV more
binding in the region of saturation due to the additional correlations from $NN$ states.
As in the case of the standard $V_{lowk}$, these $NN$ correlations beyond mean-field
also are raher weak for the density dependent $V_{lowk}(\rho)$ compared to the conventional realistic interactions and they even completely vanish at high densities.
It can be explained by the fact that only intermediate
two-particle states with momenta below the cut-off $\Lambda$ can be taken into account in the Bethe-Goldstone equation, which
in the two-particle center of mass frame
takes for an effective interaction with cut-off $\Lambda$ the form
\begin{eqnarray}
G(k',k,\epsilon_k)_{eff}=V_{eff}(k',k)+\int_0^{\Lambda} q^2 dq V_{eff}(k',q)
\nonumber \\ \frac{Q_P}{2\epsilon_k-2 \epsilon_q + i \eta} G(q,k,\epsilon_k),
\label{eq:BG}
\end{eqnarray}
where $\epsilon_i$ with $i=k,q$ are single-particle energies and $Q_P$ is the
Pauli operator. Since one has to take into account that $V_{eff}$ is designed
for a model space with relative momenta smaller than $\Lambda$, the integral in
Eq.~(\ref{eq:BG}) is restricted to momenta $q$ below the cut-off $\Lambda$.
Therefore, the $NN$ correlations become weaker at higher densities because of
the lack of phase space for these correlations in contrast to the case of
conventional realistic $NN$ interactions. In the latter case, the upper
integration limit is absent in Eq.~(\ref{eq:BG}).

\begin{figure}[!t]
\begin{center}
\includegraphics[width=0.8\textwidth] {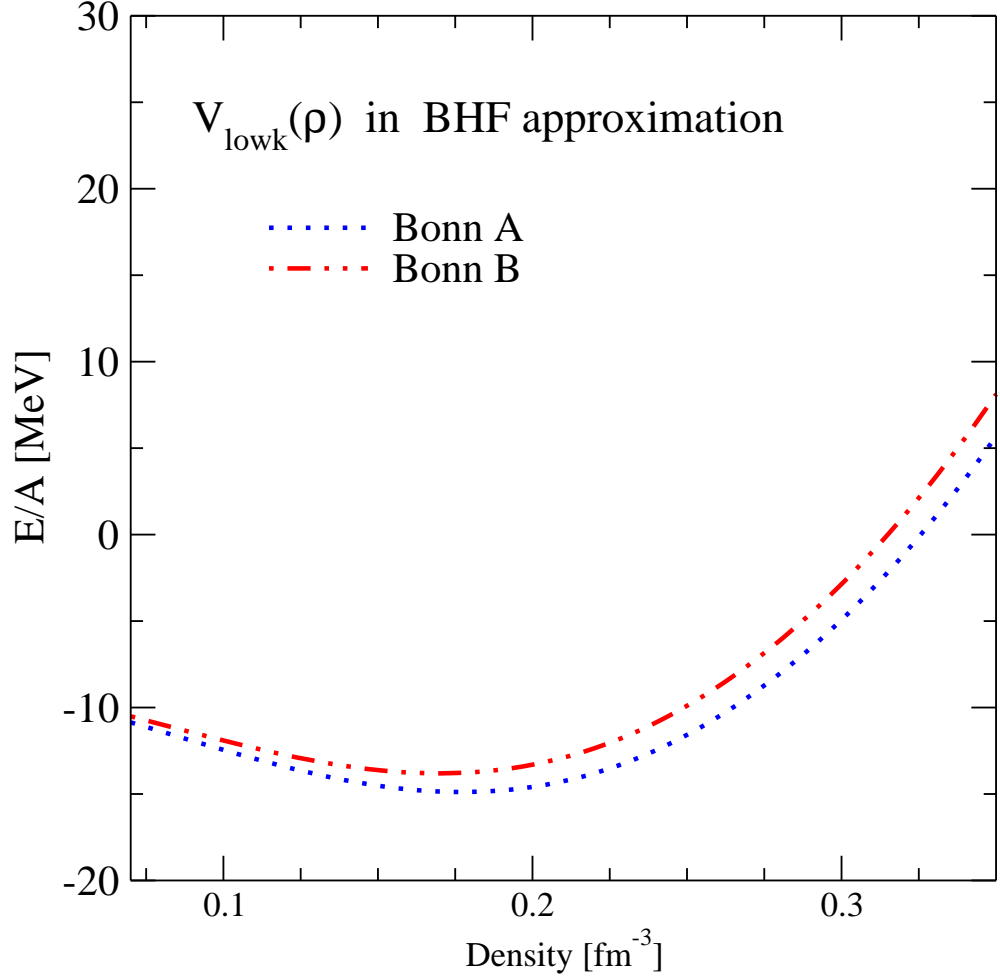}
\caption{Binding energy of isospin symmetric nuclear matter as a function of density from
BHF calculations employing the Bonn A and Bonn B potentials for the density dependent
$V_{lowk}(\rho)$ interaction. \label{fig:BHFBonnABC}}
\end{center}
\end{figure}

Another intriguing feature of the $V_{lowk}$ approach has been the finding that
the resulting low-momentum interaction is rather insensitive to the $NN$
interaction on which it is based. One obtains essentially a unique interaction
model for the low-momentum regime. The same is also true using the $V_{lowk}$ 
approach within the relativistic framework as can be seen from 
Fig.~\ref{fig:BHFBonnABC}. The energy versus density curve which are obtained in
BHF calculations using density-dependent $V_{lowk}(\rho)$ interactions based on
different meson exchange potentials are rather close to each other.  
The differences in the calculated energies are significantly smaller as those
obtained when the underlying Bonn A and Bonn B potentials are used directly
without the renormalization to $V_{lowk}$. The differences seem to be
slightly  stronger than the corresponding
model dependence in the standard $V_{lowk}$ interactions using the modern
potentials, which fit the $NN$ data with high precision. However, one must keep
in mind, that the potentials Bonn A and B did not fit the phase-shifts that
accurately and indeed a similar model dependence is also obtained applying the
renormalization scheme to these potentials in the conventional way ignoring the
relativistic effects.

In short, HF
calculations using a density dependent $V_{lowk}(\rho)$ can already lead to
reasonable results and the $NN$ correlations beyond mean-field are rather weak.
These properties of density dependent $V_{lowk}(\rho)$ give a promising prospect for studying finite nuclei in calculations which are
based on a realistic $NN$ interaction, treating correlation effects in a
perturbative manner. This should be feasible also for the description of finite
nuclei.

\section{Summary and Outlook}
We have discussed the status of relativistic nuclear many-body calculations in
nuclear matter and nuclei. These kind of calculations are based on DBHF
approaches or relativistic density functional theories such as the RMF theory.
The relativistic density functional theory is a phenomenological approach. The
parameters of these phenomenological models have been adjusted to describe
properties of isospin symmetric nuclear matter and of nuclei in the valley of
$\beta$ stability. Therefore, a problem may exist concerning the reliability for
systems at large densities and proton-neutron asymmetries. In contrast, the DBHF
approach is an \textit{ab initio} approach and the nuclear many-body problem is
treated microscopically. Hence, these DBHF approaches have a high predictive
power, since the predictions for the nuclear EoS are essentially parameter free.
A typical characteristic of both these relativistic nuclear many-body approaches
are the large scalar and vector self-energy components with a size of several
hundred MeV in contrast to nonrelativistic approaches. These large components
cancel each other to a large extent in determining the energy of the nucleons.
The large attractive scalar component of the self-energy, however, yields a
significant enhancement of the small Dirac components in the nuclear medium.

This genuine relativistic effect is described in terms of the Dirac mass, which
can only be determined from relativistic approaches. It is defined through the
scalar part of the nucleon self-energy. This Dirac mass decreases continuously
with increasing density. The DBHF calculations based on projection techniques
predict a Dirac mass splitting of $m^*_{D,n} < m^*_{D,p}$ in neutron-rich matter
in contrast to DBHF calculations based on the fit method. However,the explicit
momentum dependence has already been averaged out in the fit method, which 
makes this procedure ambiguous in isospin asymmetric nuclear matter.
Furthermore, RMF theories with the isovector $\delta$-meson included predict a
Dirac mass splitting of $m^*_{D,n} < m^*_{D,p}$. 

However, the expression of an effective nucleon mass  has also been used within
the context of nonrelativistic calculations. This nonrelativistic mass is a
measure of the nonlocality of the single particle potential. These nonlocalities
of the single particle potential can be due to nonlocalities in space or in
time. The nonrelativistic mass shows a typical peak structure around $\kf$, 
which is due the energy dependence of the nucleon self-energy or its nonlocality
in time. This peak structure is observed in relativistic as well as
nonrelativistic calculations. As a consequence of this energy dependence the 
nonrelativistic mass is not a
smooth function of the momentum in contrast to the Dirac mass. 

The Landau  mass,
i.e. the nonrelativistic mass evaluated at $k=k_F$, initially decreases with
increasing density like the Dirac mass. However, it starts to rise again at high
values of the density. The nonrelativistic mass derived from the DBHF approach
based on projection techniques in general shows a nonrelativistic mass splitting
of $m^*_{NR,n} > m^*_{NR,p}$, which is opposite to its Dirac mass splitting of
$m^*_{D,n} <m^*_{D,p}$. However, it is in agreement with results from
nonrelativistic BHF calculations. The opposite behavior between the Dirac mass
splitting and the nonrelativistic mass splitting is not surprising, since these
masses are based on completely different physical concepts.

An important bench mark for nuclear structure calculations is its success in
describing the saturation in isospin symmetric nuclear matter. Although
nonrelativistic BHF calculations were able to describe the nuclear saturation
mechanism qualitatively, they failed quantitatively. Systematic studies for a
large variety of $NN$ interactions showed that saturation points were always
allocated on a so-called Coester band in the $E/A-\rho$ plane which does not
meet the empirical region of saturation. 

The relativistic DBHF calculations do a much better job. The saturation
mechanism of these relativistic calculations is quite different compared to
nonrelativistic theories. The density dependence of Dirac spinors described in
terms of an effective Dirac mass decreasing with density leads to a density
dependence of the meson exchange interaction terms, which determines the minimum
in the energy versus density curve for nuclear matter. 

In nonrelativistic calculations the saturation point is shifted away from the 
Coester band towards the values of its relativistic counterpart only after the 
inclusion of three-body forces. Therefore, it is often argued that three-body
forces in nonrelativistic calculations are required to simulate the
 dressing of the
two-body interaction by in-medium spinors in the Dirac phenomenology of the
relativistic approaches. It should be noted, however, that there are other
sources, beside these Dirac effects, which lead to three-nucleon forces.
Examples are sub-nucleonic degrees of freedom like the medium modification of 
the meson propagators or the inclusion of excitation modes for the interacting
nucleons.

Another issue is the symmetry energy in relativistic microscopic nuclear
many-body approaches. Although the symmetry energy in nonrelativistic and
relativistic models are rather similar at a density near 0.1 fm$^{-3}$,  the
predictions are quite different at higher densities. However, this high density
behavior of the symmetry energy is essential for the description of the
structure and of the stability of neutron stars. Microscopic DBHF calculations
yield a rather stiff isospin dependence, respectively symmetry energy, at high
density, whereas non-relativistic BHF approaches give a rather soft isospin
dependence.

Many investigations have been devoted to a relativistic description of finite nuclei.
Very popular phenomenological models are relativistic density functional
theories. It is a general feature that the parameters in these purely
phenomenological theories have been adjusted to describe the saturation
properties of nuclear matter or the properties of stable nuclei located in the
valley of $\beta$ stability. Modern relativistic density functionals are able to
yield a good quantitative description of these nuclear systems. However, the
predictive power of these phenomenological interactions may be rather limited
for exotic nuclear systems such as the neutron star crust and nuclei far away
from the line of $\beta$ stability.

\textit{Ab initio} approaches such as the DBHF approach, which are
based on high precision realistic $NN$ interactions, are more ambitious.
However, full Dirac Brueckner calculations are still too complex to allow an
application to finite nuclei at present. Therefore, two different approximation
schemes have been developed: Dirac LDA and the semi-phenomenological
relativistic density functional theory.

In the Dirac LDA approach, the Dirac effects are taken into account via a kind
of local density approximation, whereas the correlations effects are taken into
account by solving the Bethe-Goldstone equation directly in the finite nucleus
under consideration. The results of these LDA Dirac calculations are generally
in better agreement with experiment than the ones of the nonrelativistic BHF
calculations, in which the medium dependence of the Dirac spinors is ignored.
The relativistic effects also improve the microscopic understanding of details
in nuclear structure like the spin orbit term. 
Further improvements may be achieved by including terms in the expansion of the
nucleon self-energy, which go beyond the BHF approximation. Such rearrangement
terms are also required to lead to a symmetry conserving approach.

In the semi-phenomenological relativistic density functional theory, the Dirac
effects are treated directly for the finite nucleus, whereas the correlation
effects are treated in a local density approximation. The results are in better
agreement with the experimental value than those from nonrelativistic BHF
approaches  and are comparable or even better than those from Dirac LDA. 

An alternative to these methods is to restrict the nuclear structure calculation
to the low-momentum components of realistic $NN$ interactions. An interesting
feature of such a low momentum potential  $V_{lowk}$ is that it is essentially
independent of the underlying realistic interaction $V$, when $V$ is fitted to
the experimental phase shifts and the cut-off $\Lambda$ is chosen around
$\Lambda$ = 2 fm$^{-1}$. However, within the nonrelativistic framework
the the $V_{lowk}$ approach fails the test of predicting nuclear matter
saturation. HF calculations employing $V_{lowk}$ do not lead to a saturation
point in symmetric nuclear matter.
 This absence of saturation is one of the major
problems of calculations employing $V_{lowk}$, since even the inclusion of
correlations beyond the HF approximation, e.g. by means of the BHF
approximation, can not cure this problem. 

The inclusion of relativistic effects to obtain a density dependent
$V_{lowk}(\rho)$ seems to be a promising prospect to solve this saturation 
problem, since HF calculations employing a density dependent $V_{lowk}(\rho)$
can already reproduce the saturation point in nuclear matter. Furthermore, this
$V_{lowk}(\rho)$ does not induce any short-range correlations into the nuclear
wave function, because the high-momentum components have been integrated out.
Mean-field calculations using $V_{lowk}(\rho)$ already lead to reasonable
results and corrections of many-body theories beyond mean-field like in the BHF
approach are weak and can be treated in a perturbative way. Therefore,
self-consistent relativistic HF calculations applying a density dependent
$V_{lowk}(\rho)$ seem to be the future for studying finite nuclei in
calculations which are based on a realistic $NN$ interaction.

\section*{Acknowledgements}
This work has been supported by a grant (Mu 705/5-2) of
the Deutsche Forschungsgemeinschaft (DFG).

\include{biblio}



\appendix


\end{document}
